\documentclass[12pt,superscriptaddress, letterpaper,aps,nofootinbib,floats,floatfix,amsmath,amssymb]{revtex4}
\usepackage{graphicx}
\usepackage{hyperref}
\usepackage{verbatim}
\begin{document}

\newcommand {\be} {\begin{equation}}
\newcommand {\ee} {\end{equation}}
\newcommand {\bea} {\begin{eqnarray}}
\newcommand {\eea} {\end{eqnarray}}
\newcommand {\eq} [1] {eq.\ (\ref{#1})}
\newcommand {\Eq} [1] {Eq.\ (\ref{#1})}
\newcommand {\eqs} [2] {eqs.\ (\ref{#1}) and (\ref{#2})}
\newcommand {\fig} [1] {fig.\ \ref{#1}}
\newcommand {\Fig} [1] {Fig.\ \ref{#1}}
\newcommand {\figs} [2] {figs.\ \ref{#1} and \ref{#2}}

\def\lsim{\mbox{\raisebox{-.6ex}{~$\stackrel{<}{\sim}$~}}}
\def\gsim{\mbox{\raisebox{-.6ex}{~$\stackrel{>}{\sim}$~}}}

\def \eps {\epsilon}
\def \veps {\varepsilon}
\def \pl {\partial}
\def \hf {{1 \over 2}}
\def \mf {\mathbf}
\def\figdir{}
\def\sss{\scriptscriptstyle}

\def\lsim{\mbox{\raisebox{-.6ex}{~$\stackrel{<}{\sim}$~}}}
\def\gsim{\mbox{\raisebox{-.6ex}{~$\stackrel{>}{\sim}$~}}}
\renewcommand{\t}{\tilde}

\title{\bf Attractive Lagrangians for Noncanonical Inflation}

\author{\normalsize Paul Franche\footnote{franchep@hep.physics.mcgill.ca}}
\affiliation{Department of Physics, McGill University\\
3600 University Street, Montr\'eal, Qu\'ebec, Canada H3A 2T8
}
\author{\normalsize Rhiannon Gwyn\footnote{rhiannon.gwyn@kcl.ac.uk}}
\affiliation{Department of Physics, King's College London\\
Strand, London, U.K.  WC2R 2LS
}
\author{\normalsize Bret~Underwood\footnote{bjwood@hep.physics.mcgill.ca}}
\author{\normalsize Alisha Wissanji\footnote{wissanji@hep.physics.mcgill.ca}}
\affiliation{Department of Physics, McGill University\\
3600 University Street, Montr\'eal, Qu\'ebec, Canada H3A 2T8
}
\date{\today}

\begin{abstract}

Treating inflation as an effective theory, we expect the effective Lagrangian to contain
higher-dimensional kinetic operators suppressed by the scale of UV physics.
When these operators are powers of the inflaton kinetic energy, the scalar field
can support a period of noncanonical inflation which is smoothly connected to the usual
slow-roll inflation.  We show how to construct noncanonical inflationary solutions to the
equations of motion for the first time, and demonstrate that noncanonical inflation is
an attractor in phase space for all small- and large-field models.
We identify some sufficient conditions on the functional form
of the Lagrangian that lead to successful noncanonical inflation
since not every Lagrangian with higher-dimensional kinetic operators can support
noncanonical inflation.  
This extends the class of known viable Lagrangians and excludes many Lagrangians which do not work.

\end{abstract}

\maketitle

\tableofcontents

\section{Introduction}

Cosmological observations of the large-scale Cosmic Microwave Background (CMB) \cite{DASI,Boomerang,WMAP} and
large-scale structure \cite{LSS,2dFGRS} are consistent with the generation of large-scale primordial
density perturbations from a period of early-universe inflation.  In the simplest models, this inflationary
period can be driven by a nearly constant energy density arising from the potential energy
of some scalar field.

There are a number of relevant energy scales in inflationary models.  At the very least
there are the Hubble scale $H$ of the inflating universe, the mass $m$ of the inflaton,
and the Planck scale $M_p$, which have a hierarchy $m \ll H \ll M_p$.  However,
all inflationary models are effective theories, only valid up to some
intermediate scale $\Lambda;\ H < \Lambda < M_p$.  When one treats inflation as an effective
theory, integrating out physics above the scale $\Lambda$ induces nonrenormalizable operators
in the effective theory,
\begin{equation}
{\mathcal L}_{eff} = {\mathcal L}_0 + \sum_{n>4} c_n \frac{{\mathcal O}_n}{\Lambda^{n-4}}\, .
\nonumber
\end{equation}
Normally, such higher-dimensional operators play a subdominant role in the effective theory
because they are suppressed by powers of the cutoff scale $\Lambda$.  However, in inflation the
self-coupling of the inflaton must be small, so higher-dimensional operators can compete with
tree-level physics.

For example, consider a canonical scalar field with a small mass, whose Lagrangian is given by
\begin{equation}
-{\mathcal L}_0 = \frac{1}{2} (\partial \phi)^2 + V_0+\frac{1}{2}m^2 \phi^2\, .
\nonumber
\end{equation}
In order for inflation to occur the mass must be sufficiently small.  The smallness
of the mass is characterized by the slow-roll parameter $\eta_{SR} = M_p^2 V''/V \approx M_p^2 m^2/V_0$, which must satisfy
$|\eta_{SR}| \ll 1$.  However, dimension-6 Planck-suppressed operators of the form
$\frac{{\mathcal O}_6}{M_p^2} \sim \frac{\langle {\mathcal O}_4\rangle}{M_p^2}\phi^2$ can lead to
corrections to the effective mass.  If $\langle {\mathcal O}_4\rangle \sim V_0$, then this higher-dimensional operator dominates the effective mass term of the inflaton and ruins the possibility of
inflation since $\eta_{SR} \sim {\mathcal O}(1)$.  This is the classic $\eta$-problem of inflation,
and demonstrates one aspect of the UV-sensitivity of inflation (see also the review by Baumann and McAllister \cite{Baumann:2009ni}
for more discussion).

Integrating out UV physics can also lead to corrections to the kinetic terms which are higher powers of derivatives.
For example, consider the two-field Lagrangian
\begin{equation}
-{\mathcal L} = \frac{1}{2} (\partial \phi)^2 + \frac{1}{2} (\partial \rho)^2 + \frac{\rho}{M} (\partial \phi)^2 + \frac{1}{2}M^2 \rho^2 + V(\phi)\, ,
\nonumber
\end{equation}
with $\phi$ our putative inflaton.  Below the mass of the $\rho$-field, i.e. for $H \ll M$, $\rho$ can be integrated
out to obtain an effective low-energy Lagrangian with higher-dimensional kinetic
terms:\footnote{See also \cite{Gelaton} for a discussion of this effect.}
\begin{equation}
-{\mathcal L}_{eff} = \frac{1}{2} (\partial \phi)^2 + \frac{(\partial \phi)^4}{M^4} + \, ...+ V(\phi)
\nonumber
\end{equation}
(where the ellipsis refers to loop corrections).  Another example is the DBI action \cite{DBI},
where integrating out $W$ bosons that couple to $\phi$ leads to an effective Lagrangian of the form
\begin{equation}
-{\mathcal L}_{DBI} = \Lambda^4 \left[\sqrt{1+\frac{(\partial \phi)^2}{\Lambda^4}}-1\right]+V(\phi)
  \approx \frac{1}{2}(\partial \phi)^2 + \frac{1}{8} \frac{(\partial \phi)^4}{\Lambda^4} + ...+V(\phi)\,.
  \nonumber
\end{equation}
It is well known that the DBI action contains novel non--slow-roll inflationary solutions
\cite{DBI,DBISky,ShanderaTye,Chen1,Chen2,DBITip}.
However, even very similar looking Lagrangians such as the tachyon action \cite{SenTachyon1,SenTachyon2,SenTachyon3,SenTachyon4},
\begin{equation}
-{\mathcal L}_{Tach} = V(\phi) \sqrt{1+\frac{(\partial \phi)^2}{\Lambda^4}},
\nonumber
\end{equation}
do not allow these novel inflationary solutions.  It is a puzzle, then, what the relevant
features of an effective Lagrangian containing higher-dimensional kinetic operators must be
so that the Lagrangian can support non--slow-roll inflation.

In this paper, we consider this question for the class of Lagrangians which are functions of the
canonical kinetic term $X \equiv -\frac{1}{2}(\partial \phi)^2$:
\begin{equation}
{\mathcal L}_{eff} = p(X,\phi)\, .
\label{eq:Leff}
\end{equation}
Lagrangians of this form
with $p(X,\phi) \rightarrow 0$ as $X\rightarrow 0$
were proposed in \cite{kinflation} as alternatives to potential energy
dominated inflation.  Those models, in which the potential energy
vanishes and the kinetic energy alone
provides the conditions for inflation, were dubbed models of {\it k-inflation}. More generally, we expect the inflationary scalar field to have
both potential and kinetic energy terms, with both terms playing important
and interesting roles during inflation.  
To avoid confusion with the specific $k$-inflation models of \cite{kinflation}, we will designate
generalized models of inflation (\ref{eq:Leff}) with both kinetic and potential energies
as models of {\it noncanonical inflation}.

The phenomenology of inflationary backgrounds from Lagrangians of the form (\ref{eq:Leff}) has
been considered before \cite{kinflationPert,NonGauss}, and these models have been studied in the Hamilton-Jacobi formalism \cite{HJ,Bean}.  However, the general inflationary solutions to the equations
of motion have not yet been found.  In Section \ref{sec:inflationsolns} we will construct the noncanonical
inflationary solutions to the equations of motion of (\ref{eq:Leff}), generalizing the well-known slow-roll results.

The noncanonical inflationary solutions trace out a trajectory in phase space.  In Section \ref{sec:attractor}
we show that these noncanonical inflationary trajectories are attractors, again generalizing known
results from the slow-roll literature.  In Section \ref{sec:GeneralAttractors} we investigate
in more detail the inflationary solutions found in Section \ref{sec:inflationsolns} and describe
some general properties of effective Lagrangians which lead to noncanonical inflationary solutions.

Certainly, (\ref{eq:Leff}) is not
the most general form for the Lagrangian.  For example, the effective Lagrangian could also
contain terms with higher derivatives
of the scalar field, e.g.~$(\partial_\mu \partial^\mu \phi)^2$.
From an effective-field-theory point of view, it is morally inconsistent to include
higher-dimensional operators like $X^2$ in the action (\ref{eq:Leff}) and neglect other
higher-derivative operators like $(\partial^2 \phi)^2$ and $(\partial^3 \phi)^2$, which have the same, or smaller, operator
dimension.  In Section \ref{sec:consistency} we will examine the consistency of keeping
only functions of $X$ in our action.  We will show that all higher-derivative terms are subleading
during inflation, being suppressed by powers of the inflationary parameters compared to the leading terms (\ref{eq:Leff}).  Thus, keeping
functions of $X$ only is an (approximately) consistent truncation of the full set of higher-dimensional
operators.

In Section \ref{sec:examples} we consider some specific Lagrangians, construct their noncanonical inflationary solutions,
and explicitly evaluate the inflationary parameters for some sample potentials, demonstrating how the procedure
for finding noncanonical inflationary solutions outlined in Section \ref{sec:inflationsolns} works in practice.
Finally, in Section \ref{sec:conclusion}, we conclude with a review of our results and a discussion of future directions.
Some technical points and auxiliary computations are relegated to the appendices.

\section{Noncanonical Inflationary Solutions}
\label{sec:inflationsolns}

We are interested in the dynamics of a single scalar field $\phi$ coupled (minimally)\footnote{A non-minimal
coupling can be written in this form after an appropriate Weyl transformation to Einstein frame \cite{Easson,Easson2}.}
to gravity with the generalized action
\begin{equation}
S = \int d^4x \sqrt{g_4}\left[\frac{M_p^2}{2}  {\mathcal R}_4 + p(X,\phi)\right]\, .
\label{eq:Action}
\end{equation}
We have written here the Lagrangian as a function of
$X\equiv -\frac{1}{2} (\partial_\mu \phi)^2$ and $\phi$.  
For homogeneous inflationary solutions we will consider the action (\ref{eq:Action}) on an FLRW
background
\begin{equation}
ds^2 = -dt^2 + a(t)^2 d\vec{x}^2\, ,
\end{equation}
with the scalar field homogeneous in space, $\phi = \phi(t)$,
so that $X\equiv \frac{1}{2} \dot{\phi}^2$ with $X > 0$ always.
The perturbations and phenomenology of inflationary backgrounds with the action (\ref{eq:Action}) have been considered
before \cite{kinflationPert,NonGauss}.  
Observables are written in terms of the {\it inflationary parameters}, generalizations of the usual slow-roll parameters familiar
from canonical inflation:
\begin{equation}
H \equiv \frac{\dot a}{a};\ \ \ \epsilon \equiv -\frac{\dot H}{H^2};\ \ \ \eta \equiv \frac{\dot \epsilon}{H \epsilon};\ \ \
  \kappa \equiv \frac{\dot c_s}{H c_s}; \ \ \ c_s^2 \equiv \left(1+2 X \frac{\partial^2 p/ \partial X^2}{\partial p/\partial X}\right)^{-1}\, .
\end{equation}
The sound speed $c_s$ is the speed at which scalar perturbations travel, and is a measure
of how far the
theory is from one with a canonical kinetic term; when $c_s$ is very different from one, the kinetic terms are strongly noncanonical.
For a canonical scalar field, we have $p(X,\phi) = X-V(\phi)$
and the inflationary parameters
reduce to the usual set of slow-roll parameters, with $c_s = 1$, and $\kappa = 0$ identically.  
The scalar power spectrum $P_k^{\zeta}$, evaluated when the perturbations
exit the sound horizon; the scalar spectral index $n_s$; and the tensor to scalar ratio $r$ are
\cite{kinflationPert} (using the notation of \cite{NonGauss})
\begin{eqnarray}
P_k^{\zeta} &= & \frac{1}{8\pi^2}\frac{H^2}{M_p^2} \left.\frac{1}{c_s \epsilon}\right|_{c_s k = a H}\,  ;\\
n_s -1 &=& -2\epsilon-\eta-\kappa \, ;\\
r &=& 16 c_s\, \epsilon \, .
\end{eqnarray}
A particularly exciting aspect of noncanonical inflationary models is that they can lead to an observable amount of non-Gaussianity in the
CMB.  The leading order behavior of the non-Gaussianity from noncanonical single field models is
of the equilateral
type \cite{NonGauss}\footnote{See \cite{Babich:2004gb} for a useful discussion of the types of non-Gaussianities.}, with amplitude
\begin{equation}
f_{NL}^{(equil)} \sim c_s^{-2}\, .
\end{equation}
Current measurements of the Gaussianity of the CMB place (shape-dependent) bounds on the amount of primordial non-Gaussianity.
For equilateral non-Gaussianities, the amplitude is constrained to
be of the order $|f_{NL}| \lsim {\mathcal O}(100)$ \cite{fnllocal,WMAP,YadavNonGauss},
which translates into a lower bound of $ c_s \gsim 0.1$ on the sound speed during the observable window of the CMB.

The equations of motion can be derived by
noticing that in the homogeneous case the scalar field can be treated
as a perfect fluid, with pressure $p = p(X,\phi)$ (hence our choice of variables) and
energy density
\begin{equation}
\rho \equiv 2 X \frac{\partial p}{\partial X} - p\, .
\label{eq:rho}
\end{equation}
The Einstein equations lead to the well-known Friedmann equations
\begin{eqnarray}
\label{eq:H}
\left(\frac{\dot a}{a}\right)^2 &\equiv & H^2 = \frac{1}{3 M_p^2} \rho; \\
\frac{\ddot a}{a} &=& -\frac{1}{6 M_p^2} (\rho+3 p)\, .
\label{eq:Friedmann2}
\end{eqnarray}
We will use an overdot to denote a derivative with respect to the
comoving time $t$.
The scalar field equations of motion can be written in several equivalent ways, depending
on the variables used. In terms of the scalar field speed $X = \dot \phi^2/2$, the equation of motion takes
the form
\begin{eqnarray}
&\dot \phi & = -\sqrt{2X} \, ;\\
&\dot X& = -2 X H \left[3-\frac{\partial p}{\partial \phi} \frac{1}{H\Pi}+ \frac{\dot p_X}{H p_X}\right] ,
\label{eq:Xeom}
\end{eqnarray}
where we have used the notation $p_X = \partial p/\partial X$, and $\Pi$ is the conjugate momentum:
\begin{eqnarray}
\label{eq:Pi}
\Pi &\equiv & \frac{\partial p}{\partial \dot \phi} = \dot \phi \frac{\partial p}{\partial X} = -\sqrt{2X} \frac{\partial p}{\partial X}.
\end{eqnarray}
In defining $\Pi$ it is assumed that $\dot \phi<0$, which is the typical case we will consider below (although it is obvious how to relax this condition).
In terms of the energy density we have
\begin{eqnarray}
\label{eq:rhoeom}
\dot \rho &=& -3 H (\rho + p) = -6 H X \frac{\partial p}{\partial X}\,.
\end{eqnarray}
We can also write the equation of motion in terms of the conjugate momentum:
\begin{eqnarray}
\dot \Pi &=& -3 H \Pi + \frac{\partial p}{\partial \phi}\, .
\label{eq:Pieom}
\end{eqnarray}

The equations of motion take equivalent forms in terms of derivatives with respect
to the number of e-folds $N_e = \int^t H d\tilde t$, which we denote by a prime, i.e.
$' \equiv d/dN_e = H^{-1} d/dt$ (where we are using the convention that the number
of e-folds is measured from the start of inflation):
\begin{eqnarray}
X' &=& -2X c_s^2 \left[3-\frac{\partial p}{\partial \phi} \frac{1}{H\Pi}+\frac{2X}{H\Pi}\frac{\partial^2 p}{\partial \phi \partial X}\right] \, ; \\
\Pi' &=& -3\left[\Pi-\frac{\partial p}{\partial \phi}\frac{1}{3H}\right] \, ;\\
\rho' &=& -3 (\rho+p)\, .
\end{eqnarray}

{}From (\ref{eq:H}) we see that for the universe to be expanding exponentially, $H$ must be approximately constant, which in turn requires the smallness of
the parameter $\epsilon$:
\begin{equation}
\epsilon \equiv -\frac{\dot H}{H^2} = -\frac{d}{dN_e} \log H \ll 1\,.
\label{eq:epsilon}
\end{equation}
Using the equation of motion (\ref{eq:rhoeom}) we can rewrite $\epsilon$ as
\begin{equation}
\epsilon = -\frac{1}{2}\frac{\dot \rho}{H\rho} = 3\frac{X}{\rho} \frac{\partial p}{\partial X}\, .
\label{eq:epsilon2}
\end{equation}
It is not sufficient that $\epsilon$ be small; in order for inflation
to last long enough, we also need the time variation of $\epsilon$ to
be small over several e-foldings, i.e.
\begin{eqnarray}
\frac{d}{dN_e} \log \epsilon = \frac{\dot \epsilon}{H\epsilon} \ll 1 \,.
\end{eqnarray}
Using (\ref{eq:epsilon2}) and $\dot \rho = -3 H \sqrt{2X}\Pi$, we can rewrite this as
\begin{equation}
\frac{\dot \epsilon}{H\epsilon} = \frac{\ddot\rho}{H\dot \rho} - \frac{\dot H}{H^2}-\frac{\dot\rho}{H\rho}
  = 4\epsilon-\left(\epsilon -\frac{\dot X}{2HX}\right)-\left(\epsilon-\frac{\dot\Pi}{H\Pi}\right)\equiv 4 \epsilon-\eta_X-\eta_\Pi \ll 1,
\label{eq:rhoddot}
\end{equation}
where we have defined for later convenience
\begin{eqnarray}
\label{eq:etaX}
\eta_X &\equiv & \epsilon-\frac{\dot X}{2 H X} = \epsilon - \frac{1}{2} \frac{d}{dN_e}\log X \, ;\\
\eta_\Pi &\equiv & \epsilon-\frac{\dot \Pi}{H \Pi} = \epsilon - \frac{d}{dN_e}\log \Pi\, .
\label{eq:etaPi}
\end{eqnarray}
The smallness of the time-variation of $\epsilon$ is commonly denoted as
$\eta \equiv \dot \epsilon/(H\epsilon)$,
as in \cite{NonGauss}.
In order for $\epsilon$ to be small for a long time, we require not only that $\epsilon \ll 1$ but also
that the condition
$|\eta| = |\dot \epsilon|/(H\epsilon) \sim |\eta_X+\eta_\Pi| \ll 1$ be satisfied.
In this paper we will consider inflationary solutions where $|\eta_X|, |\eta_\Pi| \ll 1$ individually.\footnote{In
general, one could also have $|\eta| \ll 1$ through $\eta_X \sim - \eta_\Pi$.  For a separable Lagrangian
$p(X,\phi) = q(X)-V(\phi)$ this implies $c_s^2=-1$. However in the more general case when there
is mixed $X$ and $\phi$ dependence, we can satisfy this condition with an unconstrained sound speed.}  
As we will see shortly,
when this occurs the noncanonical inflationary solutions are simple generalizations of the usual slow-roll
inflationary solutions.

With the definitions (\ref{eq:epsilon},\ref{eq:etaX},\ref{eq:etaPi}) in hand, we can
invert the equations of motion (\ref{eq:Xeom},\ref{eq:rhoeom},\ref{eq:Pieom}) to find the
general {\it noncanonical inflationary solutions} when the inflationary parameters are small
($\epsilon,|\eta_X|,|\eta_\Pi| \ll 1$):
\begin{eqnarray}
\label{eq:inflationaryrho}
p_{inf} &=& -\rho_{inf} \left(1-\frac{2}{3} \epsilon\right) \approx -\rho_{inf}\, ; \\
\Pi_{inf}(\phi) &=& \frac{\partial p}{\partial \phi} \frac{1}{3H}\left(1+\frac{\epsilon-\eta_\Pi}{3}\right)^{-1}
  \approx \frac{\partial p}{\partial \phi} \frac{1}{3H}\,.
\label{eq:PiAtrSoln}
\end{eqnarray}
The expression for $\Pi$ can either be obtained directly
from the equation of motion in the $\Pi$ variable (\ref{eq:Pieom}) or from the equation of motion in the $X$ variable
(\ref{eq:Xeom}) after using the identity $p_X'/p_X = \eta_\Pi-\eta_X$,
which is easily derived from the relation $p_X = -\Pi/\sqrt{2X}$ and the definitions of $\eta_X,\eta_\Pi$ above.

\begin{figure}[tp]
\centerline{
\includegraphics[scale=.5]{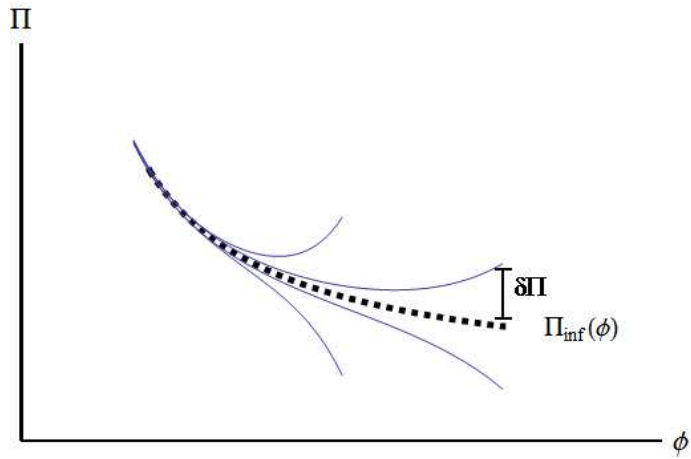}}
\caption{\small The inflationary solutions $\Pi_{inf}(\phi)$ define a trajectory in $(\phi,\Pi)$ phase
space.  This trajectory is an {\it attractor} in phase space, attracting nearby trajectories.}
\label{fig:PiInflationaryTrajectory}
\end{figure}

The expression (\ref{eq:PiAtrSoln}) defines a trajectory $\Pi_{inf}(\phi)$ [alternatively $X_{inf}(\phi)$]
in $(\phi, \Pi)$ phase space as shown in Figure \ref{fig:PiInflationaryTrajectory},
which is the inflationary trajectory as long as the inflationary
parameters $(\epsilon,\eta_X,\eta_\Pi)$ are small.  As we will see shortly, this trajectory
is also an attractor and so nearby trajectories are attracted towards it.  The inflationary parameters $\epsilon, \eta_X, \eta_\Pi$ can be
evaluated on the inflationary trajectory as simple functions of $\phi$,
\begin{eqnarray}
\label{eq:epsilonattr}
\epsilon(\phi) &=& 3\frac{X_{inf}(\phi)}{\rho(X_{inf}(\phi),\phi)} \left(\frac{\partial p}{\partial X}\right)_{X=X_{inf}(\phi) \, }\, ;\\
\label{eq:etaPiattr}
\eta_\Pi(\phi) &=& \epsilon-\frac{\dot \Pi_{inf}(\phi)}{H(X_{inf}(\phi),\phi) \Pi_{inf}(\phi)}
= \epsilon + \frac{2 X_{inf}(\phi)}{H(X_{inf}(\phi),\phi)} \frac{\frac{d}{d\phi} \Pi_{inf}(\phi)}{\Pi_{inf}(\phi)}\, ; \\
\eta_X(\phi) &=& \epsilon-\frac{\dot X_{inf}(\phi)}{2 H(X_{inf}(\phi),\phi) X_{inf}(\phi)} = \epsilon+\frac{\sqrt{2 X_{inf}(\phi)}}{2 H(X_{inf}(\phi),\phi)}\frac{\partial_\phi X_{inf}(\phi)}{X_{inf}(\phi)}\,.
\label{eq:etaXattr}
\end{eqnarray}
As we will show in Section \ref{sec:examples}, these reduce in the canonical limit to the usual slow-roll parameters: $\epsilon \rightarrow M_p^2/2 (V'/V)^2$ and
$\eta_X,\eta_\Pi \rightarrow M_p^2 V''/V$.

Let us comment on the behavior of these solutions under a local field redefinition $\phi = g(\chi)$.
Certainly, such a field redefinition leaves physical observables such as the energy density $\rho$ and
inflationary parameters $\epsilon, \eta, \kappa,c_s$ invariant.  In addition, the equation of motion
(\ref{eq:Pieom}) is invariant under such a field redefinition after the appropriate redefinition of
the conjugate momentum.  However, the secondary inflationary parameters $\eta_X,\eta_\Pi$ are clearly
{\it not} invariant under this field redefinition, transforming as
\begin{eqnarray}
\eta_X & \rightarrow & \eta_X - \frac{g'' \dot \chi}{g' H}; \label{eq:etaXtransform}\\
\eta_\Pi & \rightarrow & \eta_\Pi + \frac{g'' \dot \chi}{g' H}, \label{eq:etaPitransform}
\end{eqnarray}
so that only their sum $\eta_X + \eta_\Pi$ is invariant.
The noninvariance of $\eta_X$ and $\eta_\Pi$ under field redefinitions does not lead to any physical
inconsistencies though, since $\eta_X$ and $\eta_\Pi$ are not individually observable: only their sum
is through the invariant combination $\eta$ (\ref{eq:rhoddot}).  A similar noninvariance of the generalized
inflationary flow parameters was found in \cite{Bean}.

Our results, which use the smallness of $\eta_X$ and $\eta_\Pi$ individually, should thus be interpreted as being
true in a fixed field parametrization only, not for arbitrary parametrizations.  In particular,
take a fixed field parametrization $\phi$ in which $\epsilon, \eta_\Pi$ are small, so that (\ref{eq:PiAtrSoln}) is
the inflationary solution in that parametrization.
Substituting a field redefinition $\phi = g(\chi)$ into this inflationary solution
$\Pi^\phi_{inf} = \frac{\partial p}{\partial \phi} \left.\frac{1}{3H}\right|_{\phi=g(\chi)}$
is certainly still an inflationary solution in the new $\chi$ variable.  But this can differ
in general\footnote{Note that $\left.\frac{\partial p}{\partial \phi}\right|_{\phi=g(\chi)}\neq \frac{\partial p}{\partial \chi}$
because of additional terms picked up through the partial derivative chain rule.}
 from (\ref{eq:PiAtrSoln}) in the $\chi$ variable
precisely because $\eta_\Pi^\chi$ can be made large after a field redefinition so that the approximation
\begin{equation}
\Pi^\chi = \frac{1}{3H} \frac{\partial p}{\partial \chi} \left(1+\frac{\epsilon-\eta_\Pi^\chi}{3}\right)^{-1}
  \approx \frac{1}{3H}\frac{\partial p}{\partial \chi} \nonumber
\end{equation}
cannot be made for the $\chi$ parametrization.  Instead, the inflationary solution in the $\chi$ parametrization,
where $\eta_X$ and $\eta_\Pi$ are large, will look very nontrivial from the point of view of the equations
of motion for $\chi$.

We emphasize that this field-parametrization dependence of the inflationary solution is not troubling;
instead, it just indicates the fact that it is simpler to find the inflationary solution for some field
parametrizations than others. We are free to make a judicious choice of variable to make the behavior of the system easier to study. Viewing this from another angle, if for some particular
field parametrization we have $\epsilon$ and $\eta$ small, so that $\eta_X + \eta_\Pi$ is small, then
there exists a field redefinition such that for the new field both $\eta_X$ and $\eta_\Pi$ are
small individually.  In this new {\it preferred parametrization}, the simple inflationary solution (\ref{eq:PiAtrSoln}) applies;
thus the inflationary solution is easily found, and can easily be transformed back to the original
field definition if desired.  We will work with this {\it preferred field parametrization}
from now on without any loss of generality.

\section{Noncanonical Inflationary Attractors}
\label{sec:attractor}
\begin{figure}[htp]
\centerline{
\includegraphics[scale=.45]{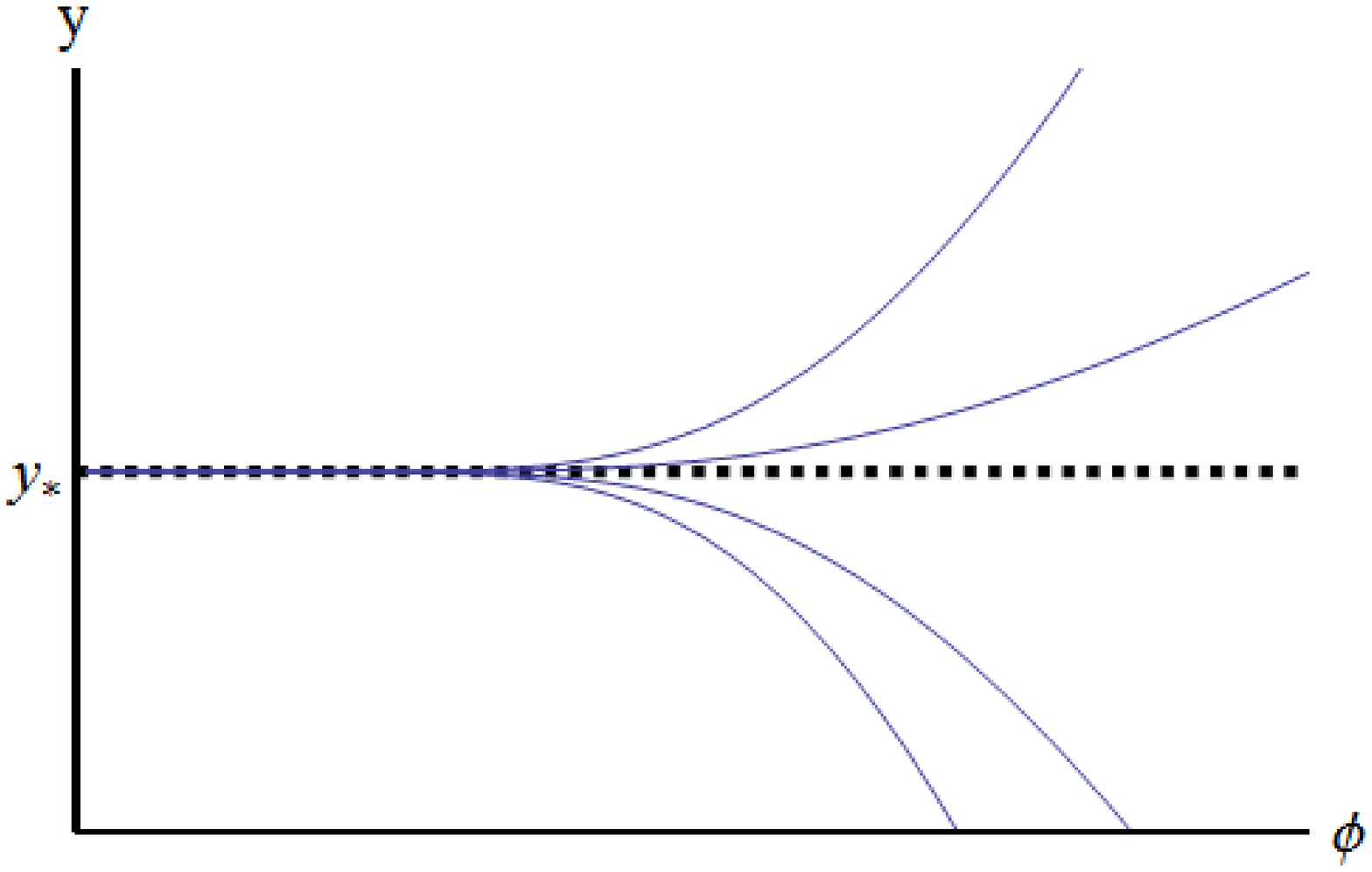}
\includegraphics[scale=.45]{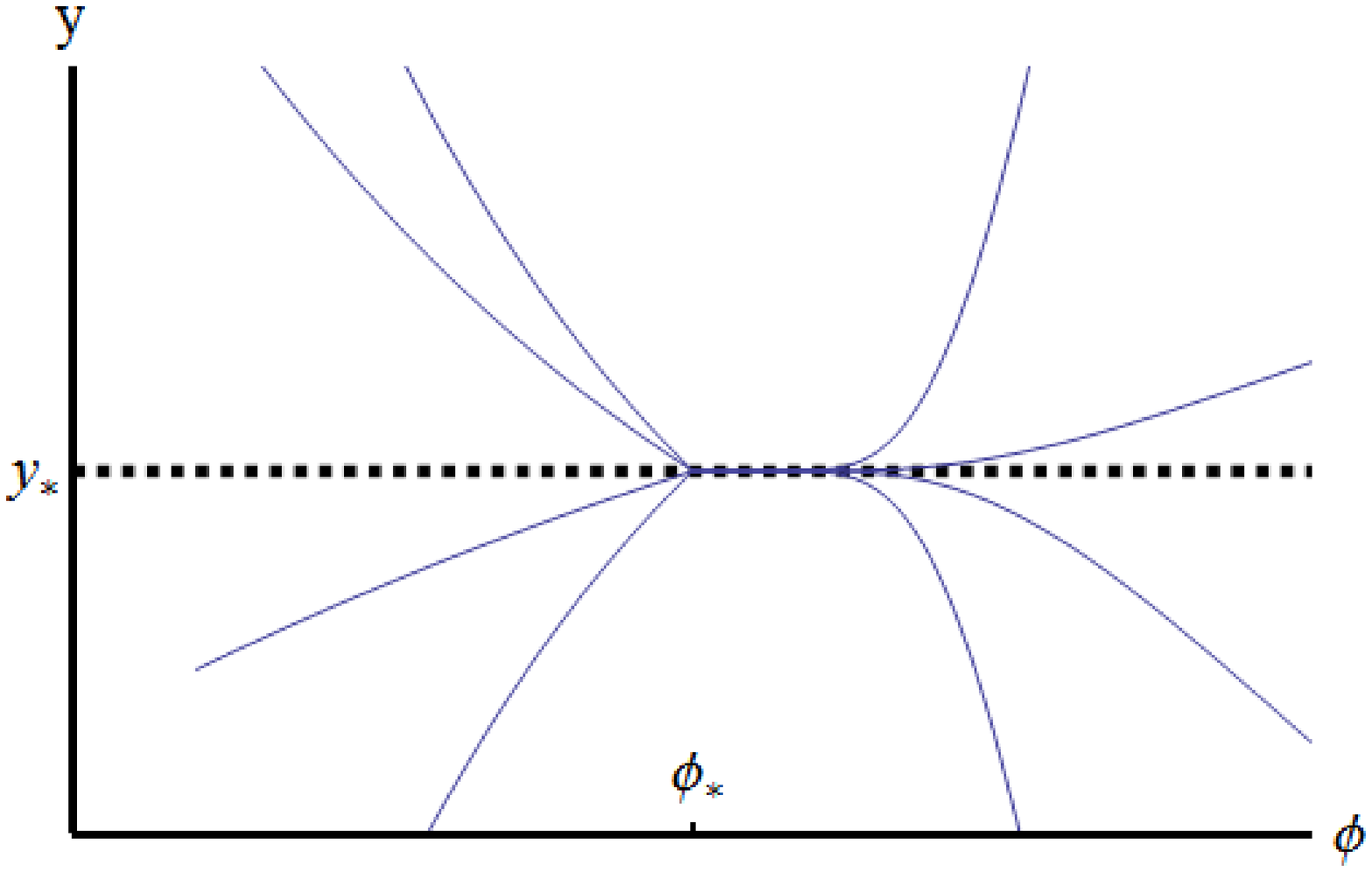}}
\caption{\small Left: The phase space for the system (\ref{eq:autoEqs}) with
$f(\phi,y) = y-y_*$ exhibits a global attractor at $y=y_*$.  Right: The phase space for $f(\phi,y) = (y-y_*)(\phi-\phi_*)$ shows the existence of an attractor
for $\phi > \phi_*$ at $y=y_*$, but this turns into a repulsor for $\phi < \phi_*$, making it a localized attractor.}
\label{fig:yAttractors}
\end{figure}

In terms of their general structure, the equations of motion
for the scalar field take the form of a pair of first order equations
(where here again a prime denotes a derivative with respect to the number
of e-folds, i.e. $' = d/dN_e$):
\begin{eqnarray}
\phi' &=& -g(\phi,y); \nonumber \\
y' &=& -f(\phi,y),
\label{eq:autoEqs}
\end{eqnarray}
where $y$ is some momentum variable
and $f,g$ are functions of
$\phi$ and $y$ determined by the equations of motion.  For example, for $y = \Pi$ we have
\begin{eqnarray}
\phi' &=& -g(\phi,\Pi); \nonumber \\
\Pi' &=& -3\left[\Pi-\frac{\partial p}{\partial \phi} \frac{1}{3H}\right]\,,
\end{eqnarray}
where $g(\phi,\Pi)$ is found by inverting the definition of $\Pi$ (\ref{eq:Pi})
(in the case where $y = X$, the equation of motion is similar).  
Differential equations
of this type are called {\it autonomous equations} and can have a number of interesting
properties, including fixed points and attractor trajectories.  For example, for a system
of equations like (\ref{eq:autoEqs}) with $f(\phi,y) = y-y_*$, where $y_*$ is a constant, the trajectory $y=y_*$ is a late-time attractor trajectory, with all trajectories approaching
$y=y_*$ at late times, as in the left-hand plot of Figure \ref{fig:yAttractors}.  
On the other hand, a system with $f(\phi,y) = (y-y_*)(\phi-\phi_*)$ for
some constant $\phi_*$ has a localized attractor at $y= y_*$ as long as $\phi > \phi_*$, but the
attractor becomes a repulsor at $\phi < \phi_*$, as in the right-hand plot of Figure \ref{fig:yAttractors}.
The term ``localized" reflects the fact that
the attractor only exists for some finite region of phase space $\phi > \phi_*$; for
$\phi < \phi_*$ the attractor becomes a repulsor.

Attractor trajectories of the system of equations (\ref{eq:autoEqs}) are
solutions $y_{att}(t)$ such that 
deviations from
this trajectory are driven to zero over time, i.e.
\begin{equation}
y= y_{att}(t) \left(1+\delta y(t)\right) \mbox{ with } \delta y(t) \rightarrow 0\,.
\label{eq:attractordeviation}
\end{equation}
More precisely, plugging (\ref{eq:attractordeviation}) into (\ref{eq:autoEqs}), we obtain an equation
for the time variation of the deviation:
\begin{eqnarray}
\frac{\delta y'}{\delta y} &=& -\left[\frac{f(\phi,y_{att}(1+\delta y))-f(\phi,y_{att})}{y_{att} \delta y}+\frac{y_{att}'}{y_{att}}\right] \\
  &\approx & -\left[\left.\frac{\partial}{\partial y} f(\phi,y)\right|_{\phi =\mbox{fixed},y=y_{att}}+\frac{y_{att}'}{y_{att}}\right]\, ,
\label{eq:yattractor}
\end{eqnarray}
where in the last line we linearized with respect to $\delta y$.
When $\delta y'/\delta y = -\beta < 0$, deviations are driven to zero exponentially fast at a rate
$\delta y \sim \delta y(0) e^{-\beta N_e}$.  
From the examples above, we see that
$\delta y'/\delta y = -1$ for the global attractor so that perturbations always decay.
For the localized attractor, $\delta y'/\delta y = -(\phi-\phi_*)$ so that perturbations decay
for $\phi > \phi_*$ and grow for $\phi < \phi_*$, in agreement with
our expectations for the existence of attractors and repulsors in the discussion above.

Before we show that inflationary solutions are attractor solutions in general, let us show that
inflation with a canonical scalar field $p(X,\phi) = X - V(\phi)$, for which $\Pi = -\sqrt{2X}$, is
an attractor in phase space.
The autonomous equations become $\phi' = \Pi/H$; $\Pi' = -3 \Pi + \partial_\phi V/H$.  Consider
perturbations about the canonical inflationary solution
\begin{equation}
\Pi_{canonical}=\Pi_{inf,canon} (1+\delta \Pi_{canonical})\, .
\end{equation}
To determine if canonical inflation is an attractor in phase space, we evaluate the equation of motion
(\ref{eq:yattractor}) for (linearized) perturbations $\delta \Pi$:
\begin{equation}
\left.\frac{\delta \Pi'}{\delta \Pi}\right|_{canonical} = -\left[3+2\epsilon_{SR}-\eta_{SR}\right] \, ,
\label{eq:PiCanattractor}
\end{equation}
where $\epsilon_{SR}$ and $\eta_{SR}$ are the usual slow-roll parameters.
When $\epsilon_{SR},|\eta_{SR}| \ll 1$, perturbations about the inflationary solution
decay as $\delta \Pi_{canonical} \sim e^{-3 N_e}$ (similar results are found using the
Hamilton-Jacobi formalism \cite{HJ}).  Thus, canonical inflation is an attractor in (homogeneous)
phase space.

Note that we have not made any assumption regarding whether we have small- or large-field inflation - the
attractor behavior holds for all types of canonical inflationary models.
This is in agreement with earlier investigations of the homogeneous initial conditions phase
space for chaotic inflation \cite{ChaoticInflation} in
\cite{Linde, Belinskyetal,Kofmanetal,PiranWilliams,Piran:1986dh,GoldwirthPiranReport,Felderetal}
and new inflation \cite{NewInflation,NewInflationLinde} in
\cite{Goldwirth,GoldwirthPiranReport}.
Note that while all models of inflation are attractors for the case of {\it homogeneous} initial
conditions, this does not imply that all models are attractors in the case of {\it inhomogeneous}
initial conditions.  

This is true more generally for any noncanonical inflationary model.\footnote{
For arguments on the existence of attractors restricted to the case when the
Lagrangian contains only kinetic terms, see \cite{kreproduction}; here we make
no assumptions about the form of the Lagrangian, so our derivation will be completely general. Further, our
analysis will not require the existence of an asymptotic $\phi \rightarrow \infty$ inflationary
attractor, as was needed in \cite{kreproduction}.}
Consider an arbitrary perturbation about the noncanonical inflationary solution (\ref{eq:PiAtrSoln}):
\begin{equation}
\Pi = \Pi_{inf} \left(1+\delta \Pi\right)\, .
\end{equation}
As before, we will evaluate the equation of motion (\ref{eq:yattractor}) for linearized perturbations
$\delta \Pi$ to determine if noncanonical inflation is an attractor.
In order to evaluate the partial derivative in (\ref{eq:yattractor})
we will use the definition of $\Pi$ to write
\begin{equation}
\left.\frac{\partial}{\partial \Pi}\right|_{\phi=\mbox{fixed}} = \frac{-c_s^2 \sqrt{2X}}{\frac{\partial p}{\partial X}}
  \left.\frac{\partial}{\partial X}\right|_{\phi=\mbox{fixed}}\, .
\nonumber
\end{equation}
Taking the partial derivative of $f(\phi,\Pi) = 3\left[\Pi-\frac{\partial p}{\partial \phi} \frac{1}{3H}\right]$
then leads to the expression (linearized in $\delta \Pi$)
\begin{equation}
\frac{\delta \Pi'}{\delta \Pi} = -3\left[1+c_s^2 \frac{\sqrt{2X}}{3H}\frac{\partial^2 p/\partial X \partial \phi}{\partial p/\partial X}
  - \frac{\partial p}{\partial \phi} \frac{\sqrt{2X}}{6\rho H} + \frac{\Pi'_{inf}}{3\Pi_{inf}}\right]_{inf}\, .
\end{equation}
Evaluting this on the inflationary solution leads to a number of simplifications, particularly after
using the definitions of the inflationary solution (\ref{eq:PiAtrSoln}), $\eta_\Pi$ (\ref{eq:etaPi}), and
$\epsilon$ (\ref{eq:epsilon}) and noticing that
\begin{equation}
\frac{\sqrt{2X}}{H}\frac{\partial^2 p/\partial X \partial \phi}{\partial p/\partial X} = \eta_\Pi - \epsilon + \frac{\epsilon-\eta_X}{c_s^2}.
\end{equation}
We soon obtain
\begin{equation}
\frac{\delta \Pi'}{\delta \Pi} = -3\left[1+\frac{1}{3}(\eta_\Pi - \epsilon)c_s^2 + \frac{1}{3}(\epsilon-\eta_X) + \frac{\epsilon}{3}+\frac{\epsilon-\eta_\Pi}{3}\right] \approx -3 + {\mathcal O}(\epsilon,\eta_X,\eta_\Pi)\, .
\label{eq:DeltaPiAttractor}
\end{equation}
As in the canonical case, we see that when the inflationary parameters $\epsilon, \eta_X,\eta_\Pi$
are small, then
perturbations decay\footnote{In principle, if the inflationary parameters are such that the right
hand side of (\ref{eq:DeltaPiAttractor}) is negative then attractive behavior may still occur,
even if the inflationary parameters are large.  This deserves further investigation.}
as $\delta \Pi \sim e^{-3 N_e}$.
Thus, we see that the noncanonical inflationary trajectories are localized attractor trajectories for linearized
perturbations.  
The attractor is localized
in the sense described above, i.e. once the inflationary parameters are no longer small the trajectory ceases
to be an attractor trajectory and thus does not necessarily continue indefinitely into the future but only exists in
some finite region of phase space.  In Appendix \ref{sec:HJ} we rederive this result in the Hamilton-Jacobi formalism
for all noncanonical inflationary models, generalizing the result of \cite{HJ}.

The derivation above only shows that noncanonical inflation is an attractor for linearized perturbations.
More generally however, we have for all $\delta \Pi$,
\begin{equation}
\frac{\delta \Pi'}{\delta \Pi} = -\left[\frac{3 \Pi_{inf}(1+\delta \Pi) - \left.\frac{\partial p}{\partial \phi} \frac{1}{3H}\right|_{\Pi=\Pi_{inf}(1+\delta \Pi)} - 3 \Pi_{inf} + \left.\frac{\partial p}{\partial \phi} \frac{1}{3H}\right|_{\Pi=\Pi_{inf}}}{\Pi_{inf}\delta \Pi} + \epsilon - \eta_{\Pi}\right]\,.
\end{equation}
For large deviations from the inflationary solution, i.e. for $\delta \Pi \gg 1$, the Hubble friction
dominates
over the driving force (proportional to $\partial p/\partial \phi$), so we have
\begin{equation}
\frac{\delta \Pi'}{\delta \Pi} \approx - \left[3+\epsilon-\eta_{\Pi}\right] + {\mathcal O}\left(\frac{1}{\delta\Pi}\right)\,.
\label{eq:infattLargePi}
\end{equation}
Thus, we see that when $\epsilon, |\eta_{\Pi}| \ll 1$, even trajectories
with arbitrarily large momentum decay as $\delta \Pi \sim e^{-3 N_e}$ and
are attracted to the inflationary attractor - inflation is a (localized)
attractor of all phase space.  This result is independent of the form of the kinetic term
as long as $\frac{\partial p}{\partial \phi}\, H^{-1}$ does not scale faster than linearly in $\Pi$ at large $\Pi$.

This does not mean that all initial conditions with arbitrarily large momentum will necessarily reach the vicinity
of the inflationary attractor in the localized region in which it exists.  Even though (\ref{eq:infattLargePi}) implies
that large momentum decays exponentially with e-folds, the Friedmann equation is dominated by kinetic energy for
large momentum.  The number of e-folds elapsed in the large momentum region will thus be small, so $\delta \Pi$ may
not decrease by a significant amount.
This distinction is particularly important for understanding the amount of fine-tuning of homogeneous initial conditions necessary for inflation, often called the {\it overshoot problem} \cite{Overshoot}.
Some investigation of the overshoot problem for DBI inflationary models has been done \cite{AttractiveBrane,Bird:2009pq}.
We plan to return to this question in the context of noncanonical inflationary models in the
future \cite{Franche:2010yj}.

\section{Conditions for Noncanonical Inflation}
\label{sec:GeneralAttractors}

The previous sections did not explicitly depend on the specific form of the scalar field Lagrangian
$p(X,\phi)$, as long as it leads to noncanonical inflation.
Unfortunately, only a few isolated examples of Lagrangians that support noncanonical inflation
exist in the literature, namely DBI inflation \cite{DBI} and k-inflation \cite{kinflation},
\begin{eqnarray}
p(X,\phi)_{DBI} &=& -\frac{1}{f(\phi)} \left[\sqrt{1-2 f(\phi) X}-1\right] - V(\phi); \\
\label{eq:DBIpgeneral}
p(X,\phi)_{k-infl} &=& f(\phi)\left(-X+\frac{X^2}{\Lambda^4}\right)\, .
\label{eq:kinf}
\end{eqnarray}
We would like to have a broader understanding of what properties noncanonical Lagrangians
should have in order to support noncanonical inflation, and to be able to write down a large class of known examples that
lead to noncanonical inflation.

We should first consider constraints on the Lagrangian such that it
makes sense as a field theory.  As noted in \cite{Bean}, we want any Lagrangian to satisfy
the null-energy condition $\rho + p = 2X \frac{\partial p}{\partial X} \geq 0$; violation of this condition
for a scalar field can lead to a number
of undesirable properties in the four-dimensional effective theory, including superluminal propagation
and ghost states.  
In addition, we want the speed of scalar perturbations to be {\it physical} (i.e. real and subluminal)
in order to avoid potential problems with acausality and UV obstructions \cite{SickTheories}
(see also \cite{Babichev:2007dw} for further discussion).  
This corresponds to a bound, $0 < c_s^2 \leq 1$, on the speed of sound
\begin{equation}
c_s^2 = \left(1+2 \frac{X \partial^2 p/\partial X^2}{\partial p/\partial X}\right)^{-1}\, ,
\label{eq:SoundSpeddDef}
\end{equation}
which leads to the conditions
\begin{eqnarray}
\frac{\partial p}{\partial X} \geq 0 && \mbox{Null Energy Condition} ;\nonumber \\
\frac{\partial^2 p}{\partial X^2} > 0 && \mbox{Physical Propagation of Perturbations}\, . \nonumber
\end{eqnarray}
For any particular Lagrangian, one must check to see if these conditions are satisfied.  It may
be possible that the violation of one or both of these conditions can be confined to some region
of phase space that is inaccessible as long as one starts initially in the physical region
of phase space, such as when the boundary between the physical and unphysical regions
is an attractor (see Figure 3 of \cite{kinflation}); this seems to be the
case with the types of models proposed in \cite{kinflation}.  
We will include an analysis of the attractor trajectories of
these types of k-inflation models in Appendix (\ref{app:kinflation}) for completeness.

We are primarily interested in models where both the potential
and noncanonical kinetic energies play important and necessary roles in inflation
(excluding models of the type (\ref{eq:kinf})).
Sometimes, as in the case of DBI inflation or the tachyon effective action, the specific form
of the Lagrangian $p(X,\phi)$ is known to all orders in $X$ and can be written in closed form.  More
generally, the Lagrangian is a
series expansion in powers of $X$,
\begin{equation}
p(X,\phi) = X + c_2(\phi) \frac{X^2}{\Lambda^4} + c_3(\phi) \frac{X^3}{\Lambda^8} + ... - V(\phi),\
\nonumber
\end{equation}
where the coefficients $c_i(\phi)$ are unknown and difficult to calculate exactly,
although some generic properties like the existence of a nonzero radius of convergence may be known.
We will thus separate our discussion of the sufficient conditions for building a successful higher-derivative
noncanonical inflationary solution into two parts, those Lagrangians which have a closed form and those which have
a series representation.  In order to have a uniform normalization of the scalar field, we
will always take the linear term in the series expansion of $p(X,\phi)$ to have unit coefficient in the small $X$ limit, i.e.~we will always
have in mind Lagrangians that reduce to
\begin{equation}
p(X,\phi)\approx X - V(\phi)
\end{equation}
at small $X$.

Before continuing further with the analysis, we will summarize our results below.
For simplicity, assume that the Lagrangian can be written in the separable form
\begin{equation}
p(X,\phi) = \Lambda^4 q\left(\frac{X}{\Lambda^4}\right) - V(\phi)\, ,
\label{eq:separablegeneral}
\end{equation}
where $q(x)$ is a function that can be either a known closed-form or some power-series representation.
This is certainly a restriction from all possible forms of the Lagrangian since it does
not allow $\phi$ dependence in the kinetic term, but it will turn out to
be a useful simplifying ansatz to work with.  We will discuss the conditions a nonseparable
ansatz should satisfy in the text.
A useful set of quantities for describing the potential $V(\phi)$ are the ``slow-roll" parameters,
\begin{eqnarray}
\epsilon_{SR} &=& \frac{M_p^2}{2} \left(\frac{V'}{V}\right)^2 ;\\
\eta_{SR} &=& M_p^2 \frac{V''}{V}\, .
\end{eqnarray}
When the potential is flat so that $\epsilon_{SR}, |\eta_{SR}| \ll 1$,
we will expect to recover canonical inflation where the generalised inflationary
parameters are identical to the slow-roll parameters, i.e. $\epsilon \rightarrow \epsilon_{SR},\ \eta_X,\eta_\Pi \rightarrow \eta_{SR}$.
Noncanonical inflation will occur when the slow-roll parameters are large, in which case
the inflationary parameters will be different from the slow-roll parameters. Nevertheless,
we will still find it useful to parameterize quantities in terms of the slow-roll parameters,
as a measure of how far the {\it potential} deviates from the slow-roll regime.

As we will see, noncanonical inflationary solutions will depend on the dimensionless variable
\begin{equation}
A \equiv \frac{V'}{3H}\frac{1}{\Lambda^2} = \left(\frac{2}{3}\epsilon_{SR} \frac{V}{\Lambda^4}\right)^{1/2}\, ,
\label{eq:Adef}
\end{equation}
where we used the assumption that the potential energy will dominate the energy density during inflation.
When $A \ll 1$, such as when the potential is very flat and $\epsilon_{SR} \ll 1$, inflation
will be canonical.  When $A \gg 1$, such as when the potential is not flat and $\epsilon_{SR} \geq 1$,
inflation will be noncanonical; in particular, the inflationary parameters will be suppressed compared to
the usual slow-roll results:
\begin{eqnarray}
A \gg 1 \Rightarrow \begin{cases}
\epsilon \approx \frac{\epsilon_{SR}}{A^m} & \cr
\eta_X \approx \frac{\epsilon_{SR}}{A^m} ,& \cr
\eta_\Pi \approx \frac{\eta_{SR}}{A^m} & \cr
\end{cases}
\end{eqnarray}
where $m$ is some number in the range $0 < m \leq 1$ which depends on the details
of the Lagrangian.  For the inflationary parameters to really be suppressed compared to
the standard slow-roll parameters when $\epsilon_{SR} \geq 1$, we need the potential to be
large in units of the effective-field-theory scale;\footnote{Strictly speaking,
we need $V/\Lambda^4> \epsilon_{SR}^{2/m-1}$.  Note also that having $V/\Lambda^4$ large
does not violate the consistency of the effective field theory treatment; as discussed in
Section \ref{sec:consistency} the effective field theory breaks down at $H\sim \Lambda$,
which corresponds to a parametrically higher scale for $V$ than is required to get noncanonical
inflation.} $V/\Lambda^4 \gg 1$.

Altogether, for a Lagrangian of the form (\ref{eq:separablegeneral}) written as a power series
\begin{equation}
p(X,\phi) = \sum_{n\geq 0} c_n \frac{X^{n+1}}{\Lambda^{4n}} - V(\phi)
\end{equation}
with a nonzero radius of convergence $X=R$, the requirements for noncanonical inflation
to be supported within the radius of convergence are:
\begin{itemize}
\item Large ``noncanonical-ness" parameter $A$: $A \gg 1$.
\item Large potential in units of the effective field theory scale: $\frac{V}{\Lambda^4} \gg 1$.
\item The derivative of the power series, $\frac{\partial p}{\partial X}$, must diverge at the radius of convergence.
\end{itemize}
The details of the power series, such as its detailed coefficients, are only important for determining whether
the derivative of the series diverges at the radius of convergence.

For a closed-form Lagrangian of the form (\ref{eq:separablegeneral}) to support noncanonical
inflation we will find it must satisfy the following properties:
\begin{itemize}
\item Large ``noncanonical-ness" parameter: $A \gg 1$.
\item Large potential in units of the effective field theory scale: $\frac{V}{\Lambda^4} \gg 1$.
\item Positively curved kinetic term: $\frac{\partial^2 q}{\partial X^2} > 0$.
\end{itemize}
Any Lagrangian that satisfies these properties will support noncanonical inflation.  Notice
that the first two properties only depend on the details of the shape and size of the potential,
while the last property, the positively curved kinetic term, is already required by our constraint
that the scalar perturbations propagate at a real and subluminal speed.

Let us now examine these cases in more detail.

\subsection{Series-Form Lagrangians}

Typically, one does not have a closed-form expression for the kinetic term of the
Lagrangian as some nontrivial function of $X$.  Instead, such terms are expected to
arise from an effective-field-theory framework in which some massive
degrees of freedom are integrated out.  In particular, recall the Lagrangian, discussed in
the Introduction, which contains a light field $\phi$ and a heavy field
$\rho$, and no terms with
higher powers of derivatives.
For energy scales smaller than the mass of the $\rho$ field, i.e. at $E \ll M$, $\rho$ can be integrated
out to obtain an effective low-energy Lagrangian with higher-dimensional kinetic terms.

In general we expect from effective field theory that our noncanonical Lagrangian will have the form of a power series:
\begin{equation}
p(X,\phi) = \sum_{n\geq 0} c_n(\phi) \frac{X^{n+1}}{\Lambda^{4n}} - V(\phi) = \Lambda^4 S(X,\phi) - V(\phi)\, .
\label{eq:general}
\end{equation}
The kinetic term series is defined to be $S(X,\phi) = \sum c_n(\phi) (X/\Lambda^4)^{n+1}$ for
convenience.  To preserve the normalization of the scalar field we will take the coefficient
of the linear term in the series to be $c_0= 1$.
This power-series form of the Lagrangian only makes sense if the
series $S(X,\phi)$ has a nonzero radius of convergence,
\begin{equation}
R \equiv \lim_{n\rightarrow \infty} \left| \frac{c_n}{c_{n+1}}\right|\, ,
\end{equation}
so that the series converges in the domain of convergence $X/\Lambda^4 \in [0,R)$.
We have in mind that $R \leq 1$; this can always be obtained by a rescaling of $\Lambda$.
It is not necessary that the series itself converge or diverge at the boundary.
Inflationary solutions for the Lagrangian (\ref{eq:general}) are
solutions to the equation (\ref{eq:PiAtrSoln}), which takes the form
\begin{equation}
\sqrt{2\frac{X}{\Lambda^4}} \sum_{n\geq 0} (n+1) c_n \left(\frac{X}{\Lambda^4}\right)^n
  = A - \sum_{n\geq 0} \Delta_n c_n \left(\frac{X}{\Lambda^4}\right)^{n+1},
\label{eq:sumAttractorEqG}
\end{equation}
where
\begin{eqnarray}
\Delta_n(\phi) &\equiv & \frac{c_n'}{c_n} \frac{\Lambda^2}{3H}\, .
\label{Deltandef}
\end{eqnarray}
The $A$ and $\Delta_n$ parameters
keep track of whether the dominant $\phi$ dependence is in the potential or kinetic
terms of the Lagrangian.

Let us consider the limit that the coefficients $c_n$ do not have strong $\phi$ dependence, i.e.
$\Delta_n \ll 1$, so that the second term on the
right-hand side of (\ref{eq:sumAttractorEqG}) is negligible.
Noncanonical inflation occurs, then, when there are solutions $X = X_{inf}(A)$ to
\begin{equation}
\sqrt{2\left(\frac{X}{\Lambda^4}\right)} \partial_X S = A(\phi)
\label{eq:sumAttractorEq}
\end{equation}
for $A \gg 1$.
The derivative of a series $\partial_X S$ has the same radius of convergence as the original
series, so (\ref{eq:sumAttractorEq}) makes sense.

The parameter that controls the nature of the inflationary solutions in (\ref{eq:sumAttractorEq})
is $A$: when $A \ll 1$ only the leading order term in the sum coming from the canonical kinetic term
in $S(X, \phi)$
contributes, and the inflationary solution reduces to the canonical
slow-roll limit with $\epsilon \rightarrow \epsilon_{SR} \ll 1$.
When $A \gg 1$, such as when the potential is no longer flat,
the higher-order terms in the sum are important and inflationary solutions become noncanonical.

Noncanonical inflationary solutions exist for a
series $S(X,\phi)$ only if (\ref{eq:sumAttractorEq}) can be solved for an arbitrary
$A \in [0,\infty)$.
If the radius of convergence $R$ is finite, then the derivative of the series $\partial_X S$
must diverge at $X/\Lambda^4=R$ in order
for solutions to exist.\footnote{Note that this does not actually require the original series itself
to diverge at $X/\Lambda^4=R$, only its derivative: even though the series and its derivative have the same
radius of convergence, they may converge or diverge
separately at the boundary. An example is the series expansion of $\sqrt{1-X/\Lambda^4}$; this series converges
at the boundary $R=1$, but the derivative of the series does not.} To see this, note that if
the series $\partial_X S$ did not diverge at $X/\Lambda^4=R$, but
instead converged to some fixed value, then there exists a maximum value the left-hand side of
(\ref{eq:sumAttractorEq}) can take; but then the solution cannot be solved for any $A$ greater
than this maximum.  Thus the series on the left-hand side must diverge at $X/\Lambda^4=R$, such that for $A \gg 1$ we have
$X_{inf}/\Lambda^4\approx R$.  

For a power series $S(X,\phi)$ that has a finite radius of convergence and a derivative
$\partial_X S$ that diverges at $R$  we have $X_{inf} \approx \Lambda^4 A^2/2$ for $A \ll 1$ and $X_{inf} \approx \Lambda^4 R$
for $A \gg 1$.
The inflationary parameters become
\begin{eqnarray}
\label{eq:epsilonseries}
\epsilon &=& 3 \frac{X_{inf} \frac{\partial p}{\partial X}|_{inf}}{\rho_{inf}} = \frac{3}{\sqrt{2}} \frac{X_{inf}^{1/2} \Pi_{inf}}{\rho_{inf}}
   = \frac{\epsilon_{SR}}{A} \left(2 \frac{X_{inf}}{\Lambda^4}\right)^{1/2}
   = \begin{cases}
   \epsilon_{SR} & A \ll 1 \cr
   \frac{\epsilon_{SR}}{A} \sqrt{2R} & A \gg 1 \cr
   \end{cases} ;\\
\label{eq:etaXseries}
\eta_X &=& \epsilon + \frac{\sqrt{2 X_{inf}}}{2H}\partial_\phi \log X_{inf}
  = \begin{cases}
  \eta_{SR} & A \ll 1\cr
  \sqrt{2R}\ \frac{\epsilon_{SR}}{A} + {\mathcal O}(\frac{\eta_{SR}}{A^3}) & A \gg 1 \cr
  \end{cases};\\
\label{eq:etaPiseries}
\eta_\Pi &=& \frac{\sqrt{2X_{inf}}}{H} \left(\frac{\partial^2 p/\partial \phi^2}{\partial p/\partial \phi}\right)
  = \begin{cases}
  \eta_{SR} & A \ll 1 \cr
  \sqrt{2R}\ \frac{\eta_{SR}}{A} & A \gg 1
  \end{cases},
\end{eqnarray}
where we used the fact that the inflationary solution has $\Pi_{inf} = \Lambda^2 A$.
Thus, we see that \em any \em series that has coefficients with $\Delta_n \ll 1$,
a finite radius of convergence, and a derivative with respect to $X$ that diverges
at this radius of convergence supports noncanonical inflation, irrespective of the rest of
the details of the power-series coefficients.

As an example, the DBI Lagrangian (with constant warp factor $f = \Lambda^{-4}$)
\begin{equation}
p(X,\phi) = \Lambda^4 \left[\sqrt{1-2\frac{X}{\Lambda^4}}-1\right]-V(\phi)
\end{equation}
has a series expansion with a radius of convergence $R=1/2$.  The series itself
converges at $X/\Lambda^4=R$, but the derivative of the series diverges at $X/\Lambda^4=R$.
Comparing the results for the inflationary parameters (\ref{eq:epsilonseries}-\ref{eq:etaPiseries})
with $R=1/2$ to the results derived later for the DBI action (\ref{epsilondbi}-\ref{etaPidbi}) in the large $A$
limit, we see that they agree perfectly.

\subsection{Closed-Form Lagrangians}

Inflationary solutions for Lagrangians of the form (\ref{eq:separablegeneral})
are found by solving (\ref{eq:PiAtrSoln}) for the inflationary solution, which
takes the form
\begin{equation}
\label{eq:separableAtr}
\sqrt{2\frac{X}{\Lambda^4}} \Lambda^4 \frac{\partial q}{\partial X} = A\, .
\end{equation}
The inflationary parameter also takes the general form
\begin{equation}
\epsilon = 3 \frac{X_{inf}}{\rho_{inf}} \Lambda^4 \frac{\partial q}{\partial X} = \frac{3}{2} \sqrt{\frac{2X_{inf}}{\Lambda^4}}\frac{\Lambda^4}{V} A
  = \frac{\epsilon_{SR}}{A} \sqrt{\frac{2X_{inf}(A)}{\Lambda^4}}\, .
\end{equation}
As long as the solution to (\ref{eq:separableAtr}) for $X_{inf}/\Lambda^4$ scales
slower than $X_{inf}/\Lambda^4 \sim A^2$ at large $A$
then we will have suppression of the inflationary parameter by powers of $A$
compared to the slow-roll result.
This happens when $\partial_X q$
grows for increasing $X$, namely when the second derivative of $q(X)$ is positive,
i.e. $q_{XX} = \frac{\partial^2 q}{\partial X^2} > 0$.
If $q_{XX}$ is zero, as in the canonical case, then we get no extra
suppression from $A$ in the inflationary parameter, as we would expect, while if
$q_{XX}$ is negative then solutions cannot, in general, be found for all $A$.
The larger the second derivative $q_{XX}$, the greater the suppression of the standard slow-roll inflationary parameter $\epsilon_{SR}$
by $A$.
It is interesting that the condition $q_{XX} > 0$ is already imposed by requiring
that the perturbations be physical, so this does not actually result in any further restrictions
on the Lagrangian.

Let us examine some examples of noncanonical Lagrangians, and show that they lead to the behavior discussed
above.
From the existence of noncanonical inflationary solutions for DBI inflation
(\ref{eq:DBIpgeneral}) \cite{DBI}, it is tempting to infer that the existence of a speed limit is a crucial feature
for the existence of noncanonical inflationary solutions.
The argument would be that for steep potentials the speed limit keeps the inflaton from moving too fast, thus
keeping it in ``slow roll".  
However, this heuristic explanation of noncanonical inflation in terms of a speed limit is not correct.
To illustrate this, consider the following ``powerlike" Lagrangian:
\begin{equation}
p(X,\phi) = \Lambda^4 \left[\left(1+\frac{2}{3} \frac{X}{\Lambda^4}\right)^{3/2}-1\right] - V(\phi)\, .
\label{eq:powerlike}
\end{equation}
This Lagrangian satisfies the physicality constraints, in that $\partial_X p, \partial^2_X p > 0$
for all values of $X$.
For $X/\Lambda^4 \ll 1$, this Lagrangian has a canonical kinetic term plus corrections,
but for $X/\Lambda^4 \gg 1$ it takes the form of a power of $X$, $p(X,\phi)\sim \Lambda^4 (X/\Lambda^4)^{3/2}-V$.  
This Lagrangian does not
imply a speed limit: $X$ can formally take any value $X\in [0,\infty)$.
The noncanonical inflationary solution can easily be found for all values of $A$;
\begin{equation}
X_{inf} = \frac{3}{4} \Lambda^4 \left[\sqrt{1+\frac{4}{3}A^2}-1\right]\, .
\end{equation}
The inflationary parameter $\epsilon$ in terms of the canonical slow-roll parameter $\epsilon_{SR}$ is
\begin{equation}
\epsilon =
  \frac{3}{2^{3/2}} \frac{\epsilon_{SR}}{A^2}\left(\sqrt{1+\frac{4}{3}A^2}-1\right)\left(1+\sqrt{1+\frac{4}{3}A^2}\right)^{1/2}
  \approx \begin{cases}
  \epsilon_{SR} & A \ll 1 \cr
  3^{1/4} \frac{\epsilon_{SR}}{A^{1/2}} & A \gg 1\,.\cr
  \end{cases}.
\end{equation}
Notice that in the large $A$ limit we obtain a suppression of the inflationary parameter
relative to the usual canonical slow-roll inflationary parameter; this implies that inflation
can occur even for a steep potential, e.g.~when $\epsilon_{SR}\sim {\mathcal O}(1)$ and $V\gg \Lambda^4$.  
Similar suppressions by powers of $A^{1/2}$ exist
for the other inflationary parameters in the large $A$ limit (the small $A$ limit gives
just the usual slow-roll results for these parameters):
\begin{eqnarray}
\eta_X \approx  \frac{3^{1/4}}{2}\frac{\epsilon_{SR}+\eta_{SR}}{A^{1/2}},\hspace{.6in} \eta_\Pi \approx  3^{1/4} \frac{\eta_{SR}}{A^{1/2}}\, .
\end{eqnarray}
No speed limit is present or necessary for this suppression.  We examine this particular example with an explicit
potential in more detail in Section \ref{sec:examples}, demonstrating that noncanonical inflationary
solutions do in fact exist.

It is straightforward to generalize this to an entire class of powerlike
Lagrangians which do not contain a speed limit but can give rise to noncanonical inflation:
\begin{eqnarray}
\label{eq:nthpower}
p(X,\phi) &=& \Lambda^4 \left[\left(1+\frac{1}{n} \frac{X}{\Lambda^4}\right)^n-1\right] - V(\phi); \\
\epsilon &\sim & \frac{\epsilon_{SR}}{A^{(2n-2)/(2n-1)}},\,\,\,\, \mbox{for } A \gg 1\, , n >1,
\end{eqnarray}
with similar expressions for the $\eta_{X,\Pi}$ inflationary
parameters.  The requirement that $n>1$ is to ensure that we satisfy the null-energy condition and physical
propagation of perturbations (although it turns out that $n < 1$ would not lead to a successful noncanonical
inflationary model anyway).
The $n\rightarrow \infty$ limit of (\ref{eq:nthpower}) takes an exponential form
\begin{equation}
\label{eq:exponential}
p(X,\phi) = \Lambda^4 \left(\sum_{n\geq 0} \frac{1}{n!} \left(\frac{X}{\Lambda^4}\right)^{n}-1\right) - V(\phi)
  = \Lambda^4 \left(e^{X/\Lambda^4}-1\right) - V(\phi)\, .
\end{equation}
Inflationary solutions are found by solving
\begin{equation}
2 \frac{X_{inf}}{\Lambda^4} e^{2 X_{inf}/\Lambda^4} = A^2\, .
\end{equation}
The solutions scale as $X_{inf}/\Lambda^4 \sim \log A$ for $A \gg 1$,
so the inflationary parameter
\begin{equation}
\epsilon \sim \epsilon_{SR} \frac{(\log A)^{1/2}}{A}
\end{equation}
is again suppressed at large $A$ compared to the usual slow roll value.
At this point, we see that an infinite number of possible Lagrangians present themselves to our study,
namely any function $q(X)$ with first and second derivative always positive, so the reader can
easily insert their favorite functional form satisfying the physicality conditions.

It is worth noting that some closed-form Lagrangians may only have noncanonical inflationary
solutions outside the regime where their power series expansion is valid.
In particular, for the powerlike class of Lagrangians,
the power series in $X$ of $\partial_X p$ all converge in their respective
domains of convergence, $X/\Lambda^4 \in [0,n]$, including the boundary.
Since we saw in the previous subsection that the derivative of the power series must diverge at the boundary
in order to support noncanonical inflationary solutions, we expect that the power series
in this case will not lead to noncanonical inflation.
Indeed, we find that the powerlike Lagrangians support noncanonical inflation for
$X/\Lambda^4 \gg n$, far outside the regime of validity for a power series expansion.
It is indeed questionable, then, whether this Lagrangian can be trusted as an effective
theory, since it is now no longer a perturbative expansion.
We would like to leave this important physical question for future analysis,
since our original purpose is to uncover the conditions a Lagrangian must have
in order to support noncanonical inflation, not whether such an
Lagrangian can or cannot be realized as an explicit effective theory.

On the contrary, the Lagrangian
\begin{equation}
p(X,\phi) = \Lambda^4 \left[1-\left(1-\frac{2}{3} \frac{X}{\Lambda^4}\right)^{3/2}\right] - V(\phi)
\label{eq:limitroot}
\end{equation}
implies a speed limit $X \leq X_{max} = \frac{3}{2} \Lambda^4$, but violates the condition
that the second derivative of the kinetic term is positive; $\partial^2 p < 0$.  Using this Lagrangian
in (\ref{eq:PiAtrSoln}),
we see that inflationary solutions can only exist when
$A < \sqrt{3/4} \Rightarrow \epsilon_{SR} \lsim {\mathcal O}(1)$,
so noncanonical inflationary solutions (those with $A\gg 1$) do not exist for this
set of Lagrangians.
Again, the speed limit is not the relevant feature of noncanonical Lagrangians that allows
inflationary solutions for steep potentials - the relevant feature is that the second derivative
of the Lagrangian with respect to $X$ is positive.

Up to this point, we have been assuming a Lagrangian which has a separable dependence on
the scalar field $\phi$ and its speed $X$, and we have seen that the only requirement
on the kinetic part of the Lagrangian is that it respect the null-energy and physical speed
of sound conditions.
To investigate the role of mixed $\phi$ and $X$ dependence further, take the simple Lagrangian
\begin{equation}
\label{eq:HD}
p(X,\phi)_{HD} = X + \frac{c_m(\phi)}{m+1}\frac{X^{m+1}}{\Lambda^{4m}} - V(\phi)
  \approx \frac{c_m(\phi)}{m+1}\frac{X^{m+1}}{\Lambda^{4m}} - V(\phi)
\end{equation}
for some $m > 0$ and $c_m > 0$.  The null-energy and physical speed of sound
conditions are satisfied for this Lagrangian for all values of $X$.
We will mostly be interested in the large $X$ and large $A$ behavior below, so we will
drop the linear term in $X$ for simplicity.
Inflationary solutions are found by solving (\ref{eq:PiAtrSoln}) as usual,
which takes the form
\begin{equation}
2^{1/2} c_m \left(\frac{X}{\Lambda^4}\right)^{m+1/2} = A + c_m \Delta_{HD} \left(\frac{X}{\Lambda^4}\right)^{m+1},
\label{eq:HDAttractorEq}
\end{equation}
where
\begin{eqnarray}
\Delta_{HD} &\equiv & -\frac{c_m' \Lambda^2}{3Hc_m}\, .
\label{DeltaHDdef}
\end{eqnarray}
Roughly, the quantities $A$ and $\Delta_{HD}$
keep track of whether the dominant $\phi$ dependence is in the
potential term or the kinetic terms.
In the limit $\Delta_{HD} \ll 1$,
the second term on the right-hand side of (\ref{eq:HDAttractorEq}) is subdominant and the inflationary solution is
\begin{equation}
X_{inf} = \frac{\Lambda^4}{2^{1/(2m+1)}c_m^{2/(2m+1)}} A_{HD}^{2/(2m+1)} \,.
\label{eq:XatrGeneral}
\end{equation}
The inflationary parameters become
\begin{eqnarray}
\epsilon &=& \epsilon_{SR} \left(\frac{2}{A^2}\right)^{m/(2m+1)}(c_m)^{-1/(2m+1)} \stackrel{m\gg 1}{\rightarrow} \sqrt{2} \frac{\epsilon_{SR}}{A}; \\
\eta_X &=& \epsilon \left(1-\frac{c_m}{2m+1}\right) + \left(\frac{2}{A^2}\right)^{m/(2m+1)}\frac{\eta_{SR}}{(2m+1) c_m^{1/(2m+1)}} \stackrel{m\gg 1}{\rightarrow} \epsilon ; \\
\eta_\Pi &=& \frac{2^{m/(2m+1)}}{c_m^{1/(2m+1)}} \frac{\eta_{SR}}{A^{2m/(2m+1)}} \stackrel{m\gg 1}{\rightarrow} \sqrt{2}\ \frac{\eta_{SR}}{A};\\
c_s^2 &=& \frac{1}{1+2m} \, ,
\end{eqnarray}
as expected from the discussion above.

Now let us consider the other limit of (\ref{eq:HDAttractorEq}), when the coefficients
of the powers of $X$ have a strong dependence on $\phi$ compared to the potential, $A \ll \Delta_{HD}$.
In this case, the inflationary solutions take the form
\begin{equation}
X_{inf}(\phi) = 2 \frac{\Lambda^4}{\Delta_{HD}^2}\, .
\end{equation}
The inflationary parameters are found to be
\begin{eqnarray}
\epsilon &=& 3\times 2^{m+1}\frac{\Lambda^4}{V} \frac{c_m}{\Delta_{HD}^{2(m+1)}}; \\
\eta_\Pi &=& \frac{2 \Lambda^2}{V^{1/2}\Delta_{HD}}\frac{\sqrt{3} M_p c_m''}{c_m'}; \\
\eta_X &=& \epsilon_{HD} - \frac{\sqrt{2X}}{H}\frac{\Delta_{HD}'}{\Delta_{HD}} = \epsilon - \eta_\Pi +3\, .
\end{eqnarray}
The difficulty with constructing an inflationary solution in this regime is due to the fact that
$\eta_X$ is generically large; thus, if an inflationary solution exists, it does not last for more
than a few e-folds.  Interesting inflationary solutions due to higher-dimensional kinetic operators
are instead found in the regime $\Delta_{HD} \ll 1$ where there is not a strong $\phi$ dependence
on the powers of the kinetic terms.

\section{Consistency of Solutions}
\label{sec:consistency}

We have been explicitly working with an effective-field-theory picture of inflation where
certain higher-dimensional operators become important, so it is important to perform
a number of checks to determine if our noncanonical inflationary solutions are internally
consistent.

First, recall that one of the  motivations for considering noncanonical Lagrangians is
that we generally expect inflation to be an EFT which has contributions
from higher-dimensional operators suppressed by powers of the UV cutoff $\Lambda$:
\begin{equation}
p(X,\phi) = \sum_{n\geq 0} c_n \frac{X^{n+1}}{\Lambda^{4n}} - V(\phi). \nonumber
\end{equation}
However, new physics at the scale $\Lambda$ should, in general, induce additional higher-dimensional operators in the EFT, such as higher powers of the curvature or its derivatives,
e.g.~$R^2/\Lambda^2, (\nabla R)^2/\Lambda^4$, or higher powers of derivatives of the scalar field,
e.g.~$\phi \, \partial^n\phi/\Lambda^{n-2}$.  It would be inconsistent, or at least rather contrived, if
only the higher-dimensional operators which are powers of $X$ appeared in the EFT, so we should check whether these additional operators are also important.

Let us first consider the higher-dimensional curvature operators.  On an inflationary background
the curvature is the Hubble scale, so the higher-dimensional curvature operators are of the order
\begin{equation}
\left\{\mbox{higher dimensional curvature operators}\right\} \sim {\mathcal O}\left(\left(\frac{H}{\Lambda}\right)^{n-4} \varepsilon^m\right),
\end{equation}
where $n$ is the dimension of the operator and $m$ is the number of derivatives in the operator.  The
quantity $\varepsilon^m$ represents an $m^{th}$ order term in the inflationary parameters, which could
be the $m^{th}$ product of the inflationary parameter $\epsilon$ or a combination
of time derivatives of the inflationary parameters, otherwise known as
the gauge-invariant inflationary flow parameters (see \cite{Bean}) generalized as
\begin{eqnarray}
\eta_X^{(i)} &=& \frac{d^i}{dN^i} \eta_X; \\
\eta_\Pi^{(i)} &=& \frac{d^i}{dN^i} \eta_\Pi\, .
\end{eqnarray}
As discussed in \cite{Bean}, the inflationary parameters $\epsilon, \eta_X^{(i)}$ and $\eta_\Pi^{(i)}$ must all
be small during inflation.  Thus we see that the higher-dimensional operators from curvature
terms are subdominant during inflation as long as $H < \Lambda$.

The condition that the Hubble rate be small compared to the UV cutoff of the EFT has import
for another physical effect related to the validity of our EFT.  Consider again the toy model
in the Introduction -- a theory of two scalar fields $\phi$ and $\rho$, with the mass $M$ of the $\rho$ field
acting as the cutoff of the resulting low-energy EFT.  We are justified in integrating out
the $\rho$ field as long as this field does not become dynamical.  However, in a quasi-de-Sitter
background quantum effects cause scalar fields to fluctuate unless the Hubble rate is smaller than
the mass of the field, $H \lsim M$.  

To be able to trust our EFT description and our truncation of the higher-dimensional curvature operators, we require $H < \Lambda$.  An important question is whether this can be satisfied consistently for noncanonical inflation.  To check this, first note that for the types of noncanonical inflation
we are largely considering here, it is the scalar field potential energy that drives inflation,
so $\rho \approx V$ up to corrections of the order of the inflationary parameters.  The Hubble rate
compared to $\Lambda$ during noncanonical inflation is then
\begin{equation}
\frac{H}{\Lambda} \approx \left(\frac{V}{\Lambda^4}\right)^{1/2} \left(\frac{\Lambda}{M_p}\right)\, .
\end{equation}
Recall that in order for noncanonical inflation to exist, the potential must be large in units of the UV scale, i.e. $V/\Lambda^4 \gg 1$. Even for the rather high GUT-scale cutoff $\Lambda\approx 10^{16} \mbox{ GeV}$,
the requirement that the Hubble rate for noncanonical inflation be small compared to $\Lambda$ translates into the inequality
\begin{equation}
1 \ll \frac{V}{\Lambda^4} < 10^4\, .
\end{equation}
There does not seem to be any obstruction to building models which
satisfy this constraint.  For UV cutoff scales smaller than the GUT scale, the EFT treatment of noncanonical
inflation should hold quite generally.  Note that unlike the case of higher-dimensional operators in the
potential, Planck-suppressed operators will not be important in the regime where the EFT is valid.

Now let us consider higher-dimensional operators containing higher powers of derivatives of the scalar field,
rewritten as terms in the Lagrangian containing
higher powers of time derivatives of $X$,
\begin{equation}
{\mathcal L} \supset p(X,\phi) + a_n(\phi,X) \frac{1}{\Lambda^m} \frac{d^m}{dt^m} X
+ \cdots, \nonumber
\end{equation}
where the ellipsis indicates other possible terms such as products of higher derivatives.
These higher-derivative operators can be ignored on the background if they are subleading compared to the
leading kinetic dependence $p(X,\phi) \sim X$.  To begin with, let us consider the operator
with one time derivative ($m=1$) and compare it with the leading $X$ dependence. Using the definition of $\eta_X$, this can be written
as
\begin{equation}
\frac{\dot X}{\Lambda X} = 2 \frac{H}{\Lambda} (\epsilon-\eta_X)\,. \nonumber
\end{equation}
As discussed above, $H < \Lambda$ is required for the EFT treatment to be
valid, so we see that on the inflationary solution $\dot X/\Lambda X \ll 1$
and we are justified in neglecting this term.
At the next order in time derivatives,
\begin{equation}
{\mathcal L} \supset \frac{\ddot X}{\Lambda^2}, \nonumber
\end{equation}
we can again use the definition of $\eta_X$ and $\eta_\Pi$ to write the ratio of this term to the leading $X$ dependence as
\begin{eqnarray}
\frac{\ddot X}{\Lambda^2 X} &=& \left(\frac{H}{\Lambda}\right)^2 \frac{\ddot X}{H^2 X} = \left(\frac{H}{\Lambda}\right)^2 \frac{\frac{d}{dN}\dot X}{HX} \nonumber \\
  &=& \left(\frac{H}{\Lambda}\right)^2 \left[-2 \epsilon(\epsilon-\eta_X)+4 (\epsilon-\eta_X)^2 + 2 \epsilon (4\epsilon-\eta_X-\eta_\Pi) - 2 \eta_X^{(1)}\right]\, .
\end{eqnarray}
In general, we find that terms in the Lagrangian with $m$ time derivatives are subleading compared
to the leading $X$ dependence by a factor
\begin{equation}
\left\{\mbox{higher derivative operators}\right\} \sim {\mathcal O}\left(\left(\frac{H}{\Lambda}\right)^{n-4} \varepsilon^m\right),
\end{equation}
where again $n$ is the dimension of the operator, $m$ is the number of time derivatives,
and $\varepsilon^m$ a term of $m^{th}$ order in the
inflationary parameters.  In models where inflation proceeds down a smooth potential, so that the tower
of inflationary flow parameters are all small, keeping just
higher powers of $X$ in the Lagrangian is a consistent truncation of the higher-dimensional operators.
Inflationary potentials which contain features like steps or bumps, such as in \cite{ChenStep,ChenStep2,AxionMonodromy},
do not automatically satisfy
these criteria and the size of these other operators, including the curvature operators, must be checked
explicitly.

Finally, let us end this section by reviewing the bounds on how noncanonical the kinetic terms
can be such that a perturbative description of inflation is still possible.
The noncanonical kinetic terms lead to couplings between the inflationary perturbations.  If these
couplings are sufficiently large then a perturbative treatment of the inflationary perturbations (as
given in \cite{kinflationPert,NonGauss}) is invalid.
In order to remain in a perturbative regime, the sound speed must satisfy the bound \cite{EFTInflation,PerturbativeInflation,Shandera:2008ai},
\begin{equation}
c_s > \left(\frac{H}{M_p}\right)^{2/5} \frac{1}{\epsilon^{1/5}}\, .
\label{eq:csbound}
\end{equation}
Alternatively, we can view the perturbations as becoming strongly coupled at the scale \cite{EFTInflation}
\begin{equation}
\Lambda_{strong} \sim \left(M_p H\right)^{1/2} \epsilon^{1/4} c_s^{5/4}
\end{equation}
when $c_s \ll 1$.  Demanding $H < \Lambda_{strong}$ leads to (\ref{eq:csbound}).

For noncanonical inflation, the sound speed $c_s$ depends on the parameter $A$ controlling the noncanonical
nature of inflation, and so (\ref{eq:csbound}) can be turned into a bound on $A$ for specific models.
For example, for DBI inflation we have $c_s \sim A^{-1}$ (\ref{eq:csDBI}), so this places an upper bound on $A$ for
DBI inflation, $A < (M_p/H)^{2/5} \epsilon^{1/5}$, which places stringent limits on the perturbative
regime of DBI inflation.  Alternatively, for the powerlike Lagrangian (\ref{eq:powerlike})
we have $c_s > 1/\sqrt{2}$ for all $A$, so (\ref{eq:csbound}) does not restrict the allowed values of $A$ at all.
Instead, we can turn this into an upper bound on the Hubble rate, $H/M_p < \epsilon^{1/2}$.  
These bounds on the perturbative regime of noncanonical inflation seem to favor low-scale inflation and limit
the degree of ``noncanonical-ness" of inflation.

\section{Examples}
\label{sec:examples}

In this section, we will explore the noncanonical inflationary solutions constructed above for several popular
higher-derivative Lagrangians.  

\begin{figure}[tp]
\centerline{
\includegraphics[scale=.5]{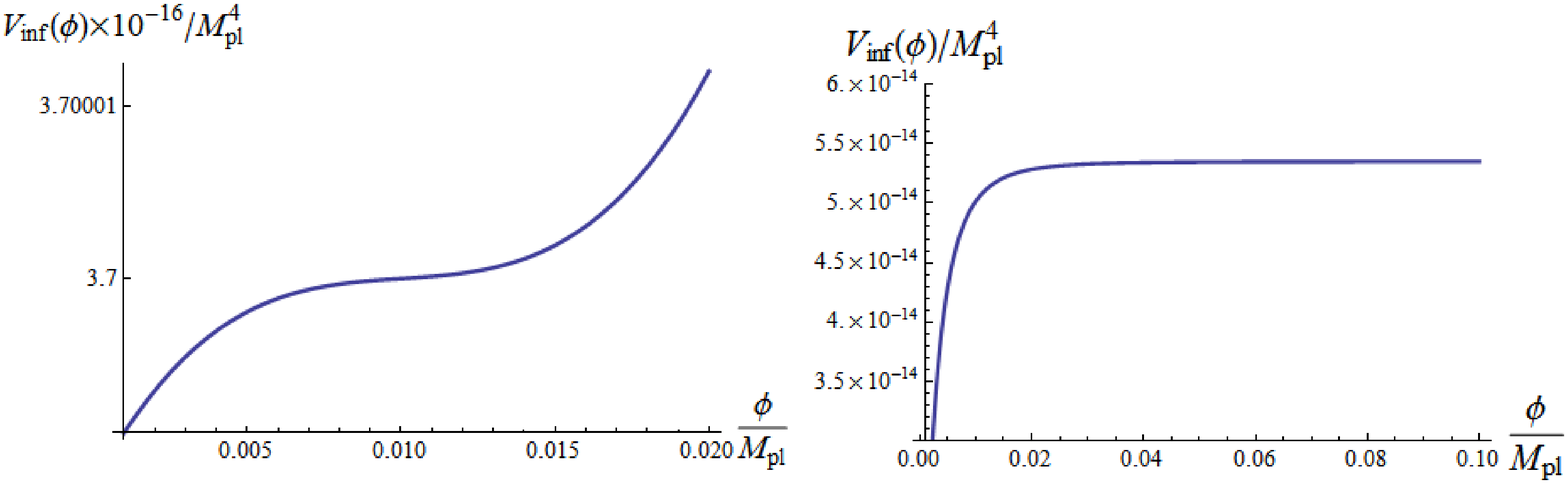}}
\caption{\small Left: An inflection-point-type potential (\ref{eq:inflectionpt}).  Right:
A Coulomb-type potential (\ref{eq:coulomb}). The values of the parameters are given in (\ref{eq:potparams}).}
\label{fig:potentials}
\end{figure}

Throughout this section, we will be working with Lagrangians of the form
\begin{equation}
p(X,\phi) = q(X,\phi) - V(\phi),
\end{equation}
such that $q(X,\phi) \approx X$ at small $X$ (with one exception, the tachyon effective action).
In order to illustrate the noncanonical nature of the inflationary solutions, we will study
the same two potentials for each Lagrangian below, an ``inflection-point" type potential
and a ``Coulomb" type potential, shown in Figure \ref{fig:potentials}.
\begin{eqnarray}
\label{eq:inflectionpt}
V(\phi)_{inflection} &=& V_0 + \lambda (\phi-\phi_0) + \beta (\phi-\phi_0)^3; \\
V(\phi)_{coulomb} & =& V_0 - \frac{T}{(\phi+\phi_0)^n}\, .
\label{eq:coulomb}
\end{eqnarray}
Both of these potentials have some basis in brane- and string-inflation model building.
Inflection-point potentials have arisen in models of D-brane inflation in warped throats
\cite{Delicate,ExplicitDbrane,HolographicSystematics}
and in closed string models \cite{Accidental,Badziak:2008gv}.
Coulomb potentials have appeared as potentials for D-brane inflation
\cite{DvaliTye} and its embeddings in warped geometries \cite{KKLMMT}.

We have chosen the parameters such that {\it canonical} inflation on these potentials gives rise to $60$ or
more e-folds of inflation,
\begin{eqnarray}
\mbox{Inflection Point:}&& \ \left(V_0 = 3.7\times 10^{-16},\lambda = 1.13\times 10^{-20},\beta = 1.09\times 10^{-15},\phi_0 = 0.01\right) \nonumber; \\
\mbox{Coulomb:}&& \ \left(V_0 = 5.35\times 10^{-14}, T=\phi_0^4 V_0, n=4, \phi_0 = 0.01\right)\,,
\label{eq:potparams}
\end{eqnarray}
and so that the normalization of the power spectrum and the spectral index agree with
observations \cite{WMAP}, i.e. $P_\zeta = 2.41\times 10^{-9},\ n_s = 0.961$.

\subsection{Canonical Inflation}

Let us see how the prescription (\ref{eq:PiAtrSoln}) for computing the
inflationary attractor solutions works by first computing the well-known attractors
for the case when the Lagrangian consists of a canonical kinetic term and potential:
\begin{equation}
p_{canon} = X-V(\phi)\,.
\label{eq:canon}
\end{equation}
It is straightforward to find that
\begin{eqnarray}
\Pi &=& -\sqrt{2X} \\
\Rightarrow \frac{\dot\Pi}{H \Pi} &=& \frac{1}{2} \frac{\dot X}{HX}\,\,\, \Rightarrow \eta_{\Pi} = \eta_X\, .
\end{eqnarray}
The energy density is simply $\rho = X+V$, so solving (\ref{eq:PiAtrSoln}) for $X$ gives the usual solution:
\begin{eqnarray}
X_{inf}(\phi) &=& \frac{1}{2} V(\phi) \left[\sqrt{1+\frac{2}{3} M_p^2 \left(\frac{V'}{V}\right)^2}-1\right]
  \approx \frac{M_p^2}{6} \frac{(V')^2}{V}\, ;
\label{eq:XatrCanon} \\
\Pi_{inf}(\phi) &\approx & -\frac{M_p V'}{\sqrt{3} V^{1/2}}\, .
\end{eqnarray}
The approximation is made to leading order in $\epsilon$.
It is straightforward to evaluate the inflationary parameters (\ref{eq:epsilonattr}-\ref{eq:etaXattr})
using the solution (\ref{eq:XatrCanon}).
The inflationary parameters reduce to the usual ``slow-roll" parameters in terms of the derivatives of the potential;
\begin{eqnarray}
\epsilon &=& \frac{M_p^2}{2} \left(\frac{V'}{V}\right)^2, \\
\eta_X &=& \eta_\Pi = M_p^2 \frac{V''}{V}\, .
\end{eqnarray}

\begin{figure}[tp]
\centerline{\includegraphics[scale=.43]{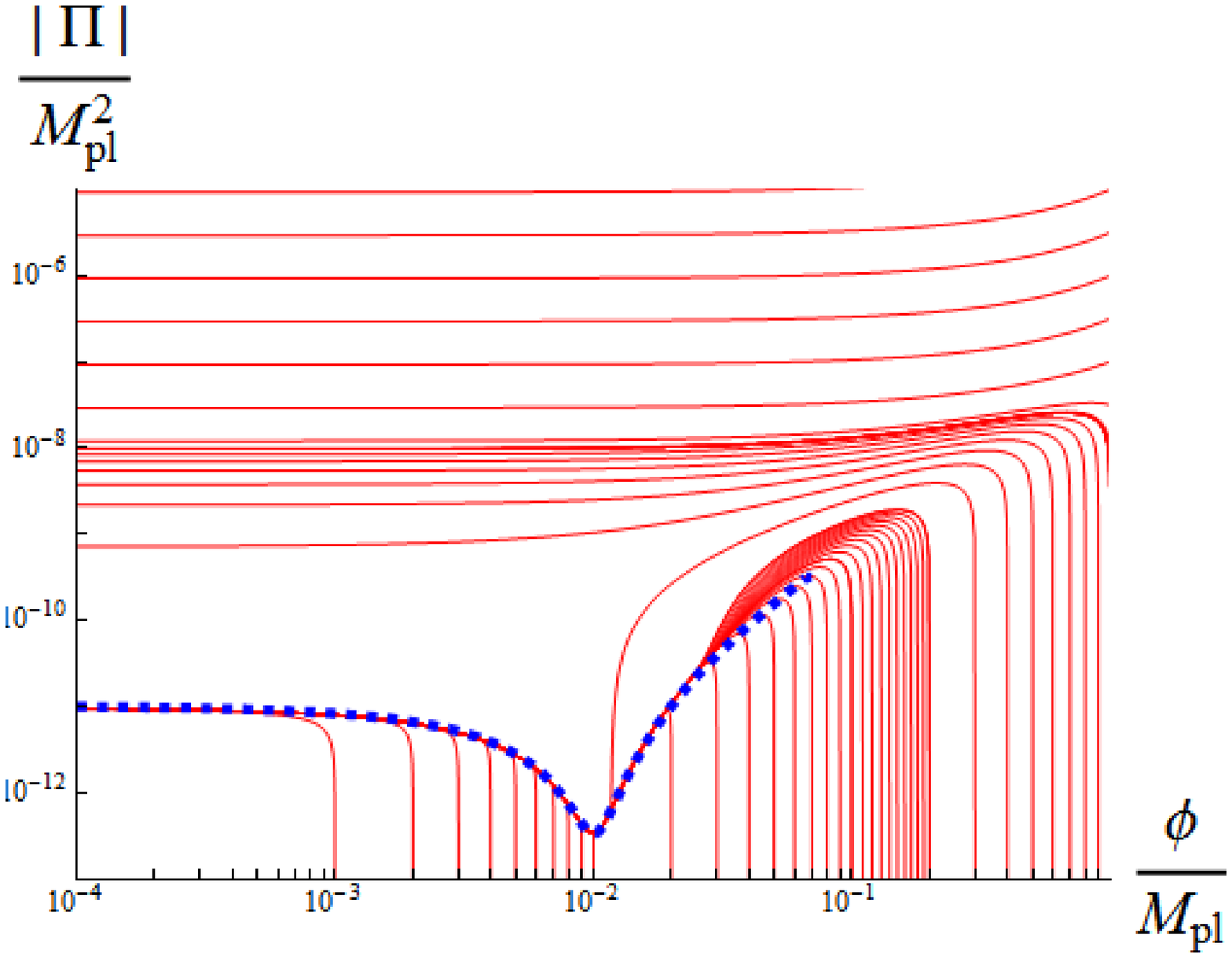} \includegraphics[scale=.48]{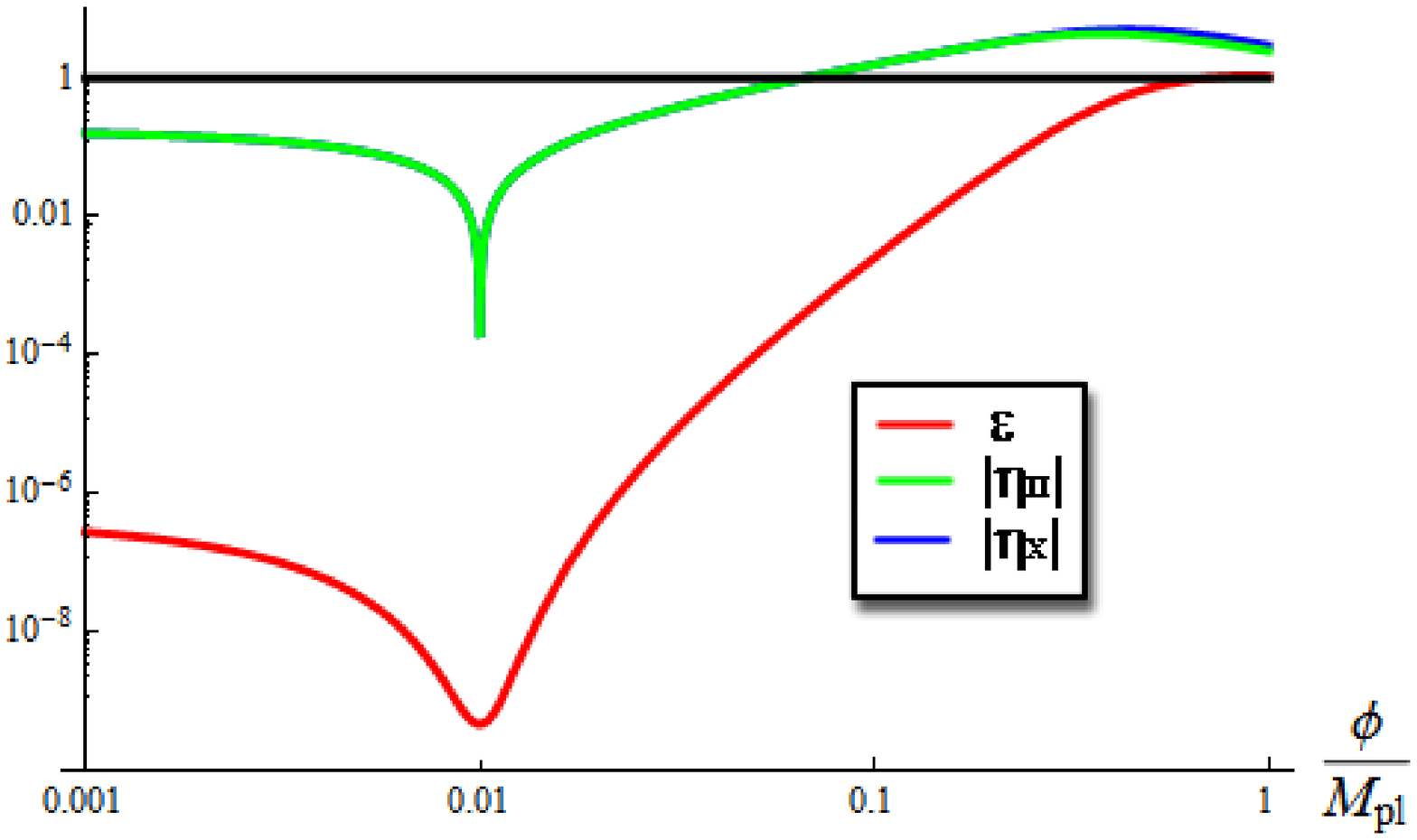}}
\caption{\small Left: The $(\phi,|\Pi|)$ phase-space diagram for a scalar field with a canonical kinetic term
and inflection-point potential (\ref{eq:inflectionpt}).  The thick dashed line is the inflationary attractor trajectory
and thin solid lines are exact numerical solutions to the equations of motion.
Right: The inflationary parameters $\epsilon,|\eta_X|$, and $|\eta_\Pi|$ are shown as functions of $\phi$. The solid horizontal (black)
line denotes the value 1.}
\label{fig:CanonPhasePlot}
\end{figure}

\begin{figure}[htp]
\centerline{\includegraphics[scale=.40]{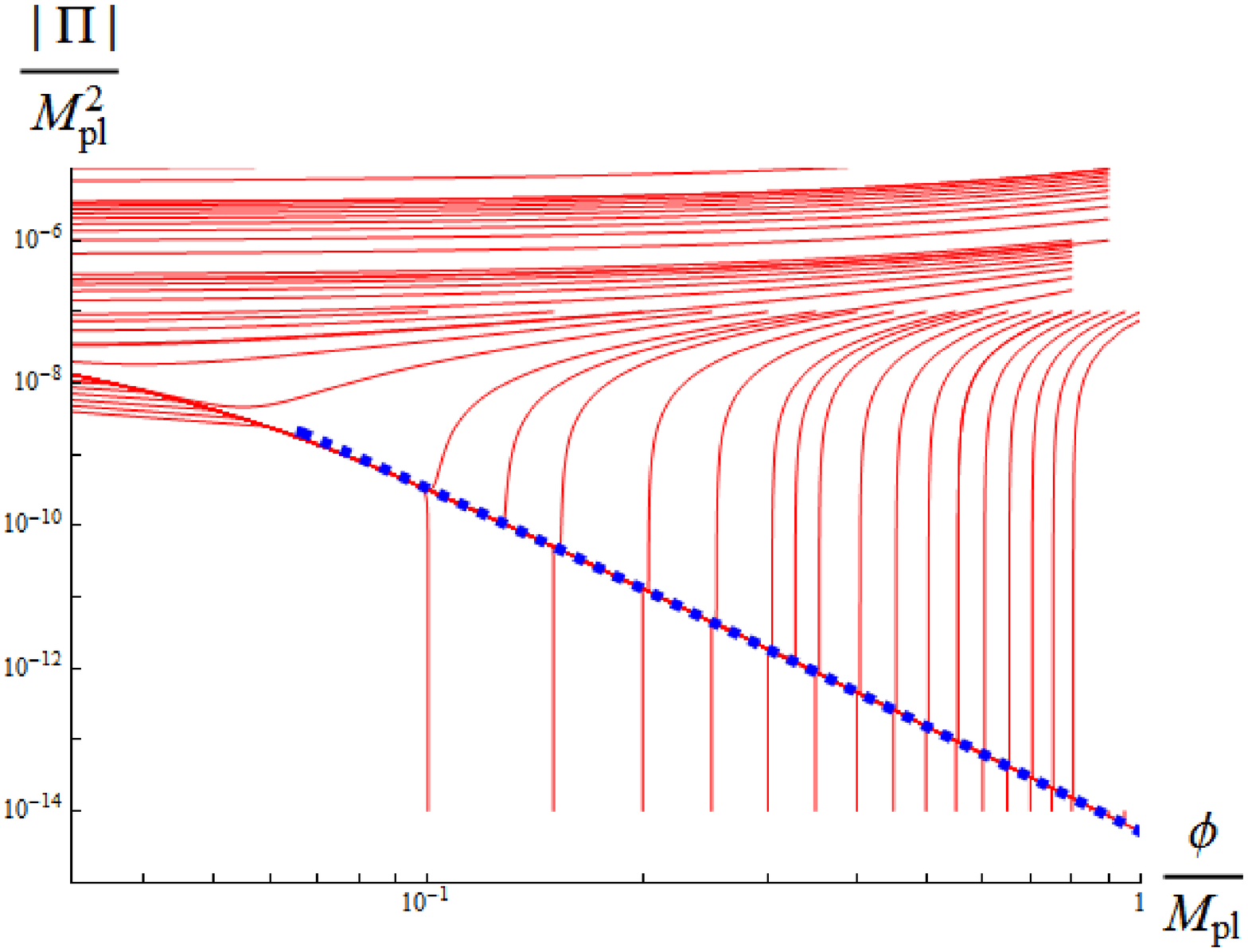} \includegraphics[scale=.48]{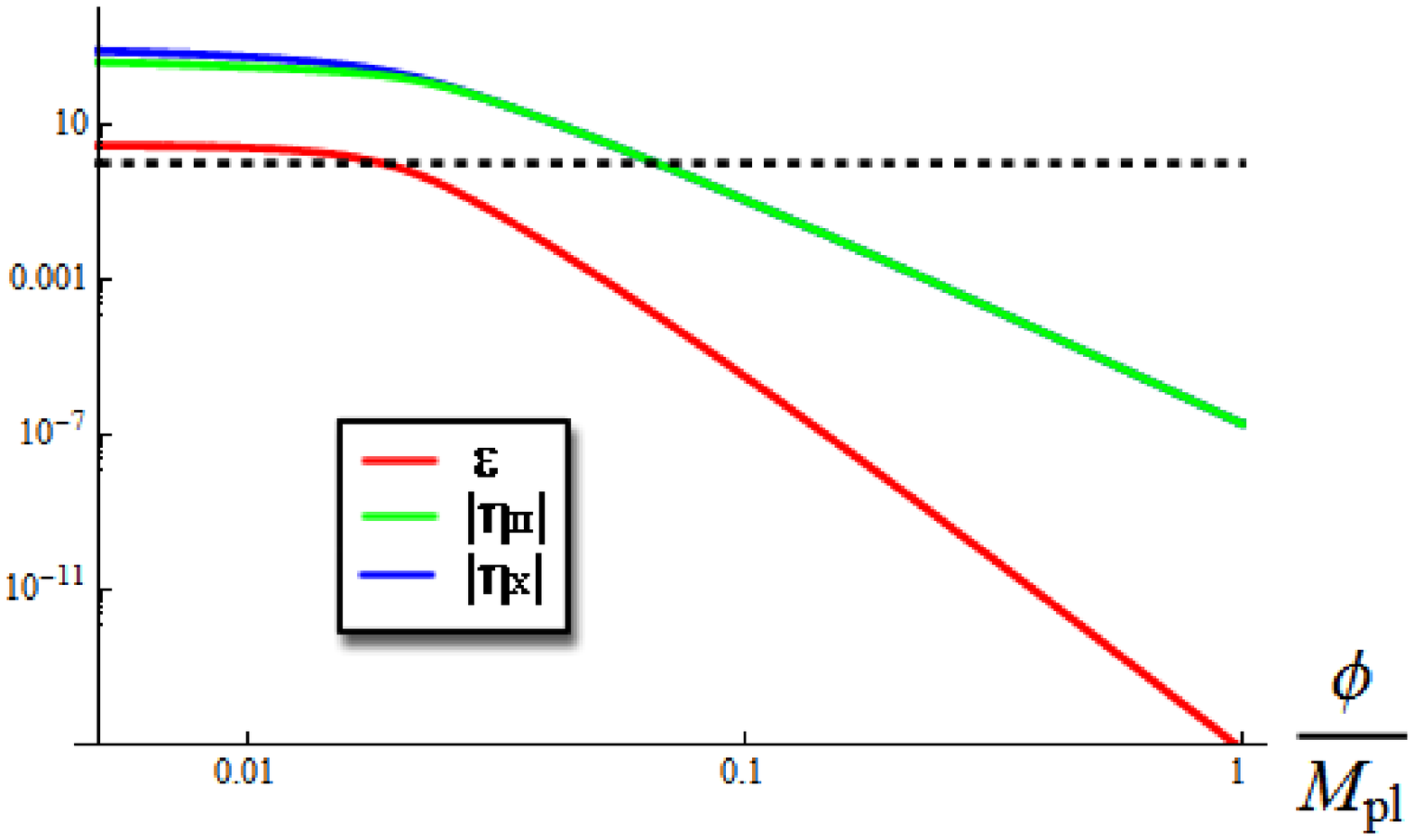}}
\caption{\small Left: The $(\phi,|\Pi|)$ phase-space diagram for a scalar field with a canonical kinetic term
and Coulomb potential (\ref{eq:coulomb}). The thick dashed line is the inflationary attractor trajectory
and thin solid lines are exact numerical solutions to the equations of motion.
Right: The inflationary parameters $\epsilon,|\eta_X|$, and $|\eta_\Pi|$ are shown as functions of $\phi$. The solid horizontal (black)
line denotes the value 1.}
\label{fig:CanonPhasePlot_Coulomb}
\end{figure}

The canonical inflationary attractor solution in phase space for the inflection-point potential is shown as a thick dashed
line in the left-hand plot of Figure \ref{fig:CanonPhasePlot}, along with some sample trajectories for initial
conditions outside of the inflationary regime.  The inflationary parameters for a canonical kinetic term
and inflection-point potential are shown on the right-hand side of Figure \ref{fig:CanonPhasePlot}, where we
see that $\eta_X = \eta_\Pi$, as expected by our analysis.
The same variables are shown in Figure \ref{fig:CanonPhasePlot_Coulomb} for the Coulomb-like potential (\ref{eq:coulomb}).

\subsection{DBI Inflation}

One popular single-field Lagrangian which contains higher-dimensional derivative operators is
the Dirac-Born-Infeld (DBI) Lagrangian,
\begin{equation}
p(X,\phi) = -\frac{1}{f(\phi)} \left(\sqrt{1-2 f(\phi) X}-1\right) - V(\phi)\, .
\label{eq:DBIP}
\end{equation}
This type of Lagrangian appears when describing the motion of $D3$-branes in a warped
product space of the form $\mathbb{R}^{3,1}\times {\mathcal M}_6$
(see \cite{DBI,DBISky,Meng,ShanderaTye,AttractiveBrane} for some discussion of DBI inflation and
DBI attractors); here we will take (\ref{eq:DBIP}) as a phenomenological model.
The reality of the Lagrangian implies a speed limit for the scalar field:
\begin{equation}
X_{max} = \frac{1}{2 f(\phi)}\, .
\end{equation}
The function $f(\phi)$, called the ``warp factor" since it arises when a D3-brane is embedded
in a warped space, can be viewed as a field-dependent UV cutoff, i.e. $f(\phi) = \Lambda_{eff}^{-4}$,
arising from integrating out $W$ bosons with a field-dependent mass.
The conjugate momentum for the DBI Lagrangian (\ref{eq:DBIP}) is
\begin{equation}
\Pi_{DBI} = -\frac{\sqrt{2X}}{\sqrt{1-2 f(\phi) X}}\, .
\end{equation}

In order to find the inflationary solution, we must solve (\ref{eq:PiAtrSoln}) for $X_{inf}(\phi)$.  The full expression
to be solved is
\begin{equation}
\sqrt{3}\chi \sqrt{\frac{1}{\sqrt{1-\chi^2}}-1+f(\phi) V(\phi)} = M_p V' f \sqrt{1-\chi^2} - \frac{M_pf'}{f} \left(1-\frac{1}{2}\chi^2-(1-\chi^2)^{1/2}\right)\,,
\end{equation}
where we used the variable $\chi \equiv \sqrt{2Xf(\phi)}$ to simplify the expression.
The general solutions for $\chi(\phi)$ to this algebraic equation are difficult to find analytically.  
However, we can simplify the expression
using variables analogous to those defined above (in (\ref{eq:Adef}), (\ref{Deltandef}) and (\ref{DeltaHDdef})) and in \cite{AttractiveBrane}:
\begin{eqnarray}
A_{DBI}(\phi) &\equiv & \frac{V'(\phi) f^{1/2}(\phi)}{3 H(\phi)}; \\
\Delta_{DBI}(\phi) &\equiv & \frac{f'(\phi)}{3H(\phi) f^{3/2}(\phi)},
\label{eq:Deltadef}
\end{eqnarray}
where a prime $'$ denotes a derivative with respect to $\phi$.
We then find an expression for the inflationary
solution $\chi_{inf}(\phi)$ which looks similar to Eq.(3.27) of \cite{AttractiveBrane} (which determined the ``fixed
point" solutions, in the terminology of that work):
\begin{eqnarray}
&& V' f^{1/2} (1-\chi^2)^{1/2}-3H\chi-\frac{f'}{f^{3/2}}(1-\frac{1}{2}\chi^2-(1-\chi^2)^{1/2}) = 0 \nonumber \\
&&\Rightarrow A_{DBI} (1-\chi^2)^{1/2}-\chi-\Delta_{DBI} \left(1-\frac{1}{2}\chi^2-(1-\chi^2)^{1/2}\right) = 0.
\label{eq:chifixedpt}
\end{eqnarray}

The differences between the expression (\ref{eq:chifixedpt}) determining the attractor solution and the corresponding
expression Eq.(3.27) of \cite{AttractiveBrane} are found to be only in the terms multiplying $\Delta_{DBI}$.  These differences
can be understood, since \cite{AttractiveBrane} was looking for attractor solutions of the form $\dot \chi \approx 0$,
while the solutions of (\ref{eq:chifixedpt}) are attractor solutions of the form $\dot \Pi \approx 0$.  Since
the two variables $\chi$ and $\Pi$ are related by $\Pi = -f^{-1/2}\chi/\sqrt{1-\chi^2}$, it is not surprising that we find
some differences in the corresponding attractor solutions. Despite the overlap in the solutions, the approach given here [using
(\ref{eq:PiAtrSoln}) or (\ref{eq:chifixedpt})] is more general than that presented in \cite{AttractiveBrane} because
it can be applied to other Lagrangians beyond DBI.

We show in Appendix \ref{sec:DBIDelta} that in order for DBI inflationary solutions to (\ref{eq:chifixedpt})
to exist, we require $|\Delta_{DBI}| \ll 1$.  In this limit, the inflationary solutions are
\begin{eqnarray}
\Pi_{inf}^{DBI}(\phi) &=& -\frac{M_p V'}{\sqrt{3 V}}; \\
X_{inf}^{DBI}(\phi) &=& \frac{ A_{DBI}^2 }{2 f(\phi)(1+A_{DBI}^2)}\,. \label{eq:XDBI}
\end{eqnarray}
The sound speed evaluated on the inflationary solution is
\begin{equation}
c_s^2 = \frac{1}{1+A_{DBI}^2},
\label{eq:csDBI}
\end{equation}
so $A_{DBI} \gg 1$ implies the noncanonical limit $c_s \ll 1$, while $A_{DBI} \ll 1$ implies the canonical limit $c_s \sim 1$.
The inflationary parameters (\ref{eq:epsilonattr}-\ref{eq:etaXattr}) become simple functions of $\phi$ on the inflationary
solution:
\begin{eqnarray}
\epsilon_{DBI} &=& \frac{\frac{3}{2}A_{DBI}^2}{\left(1+A_{DBI}^2 \right)}\frac{1}{1+\frac{f(\phi)V(\phi)-1}{(1+A_{DBI}^2)^{1/2}}} \approx
\begin{cases}
\frac{3}{2} \frac{A_{DBI}^2}{fV} = \epsilon_{SR} & A_{DBI} \ll 1 \mbox{ ``Slow Roll"} \cr
\frac{3}{2} \frac{A_{DBI}}{fV} = \frac{\epsilon_{SR}}{A_{DBI}} & A_{DBI} \gg 1 \mbox{ ``DBI"}\cr
\end{cases};\label{epsilondbi}\\
(\eta_X)_{DBI} &=& \epsilon_{DBI} + \frac{\eta_{SR}-\epsilon_{SR}}{(1+A^2)^{3/2}} \approx
\begin{cases}
\eta_{SR} & A_{DBI} \ll 1 \mbox{``Slow Roll"} \cr
\frac{\epsilon_{SR}}{A_{DBI}} & A_{DBI} \gg 1 \mbox{ ``DBI"}\cr
\end{cases} ;\label{etaXdbi} \\
(\eta_\Pi)_{DBI} &=& \frac{\eta_{SR}}{(1+A_{DBI}^2)^{1/2}} \approx
\begin{cases}
\eta_{SR} & A_{DBI} \ll 1 \mbox{``Slow Roll"} \cr
\frac{\eta_{SR}}{A_{DBI}} & A_{DBI} \gg 1 \mbox{ ``DBI"}\cr
\end{cases};\label{etaPidbi}
\end{eqnarray}
where we have used the fact that $fV$ must be much greater than $A_{DBI}$ to simplify the large $A_{DBI}$ or DBI limit of $\epsilon_{DBI}$.
Notice that in the large $A_{DBI}$ limit the inflationary parameters are suppressed compared to their usual
slow-roll values by a power of $A_{DBI} \gg 1$; this is the effect, noticed previously, that a DBI kinetic
term can support inflation even on a steep potential.

\begin{figure}[tp]
\centerline{\includegraphics[scale=.43]{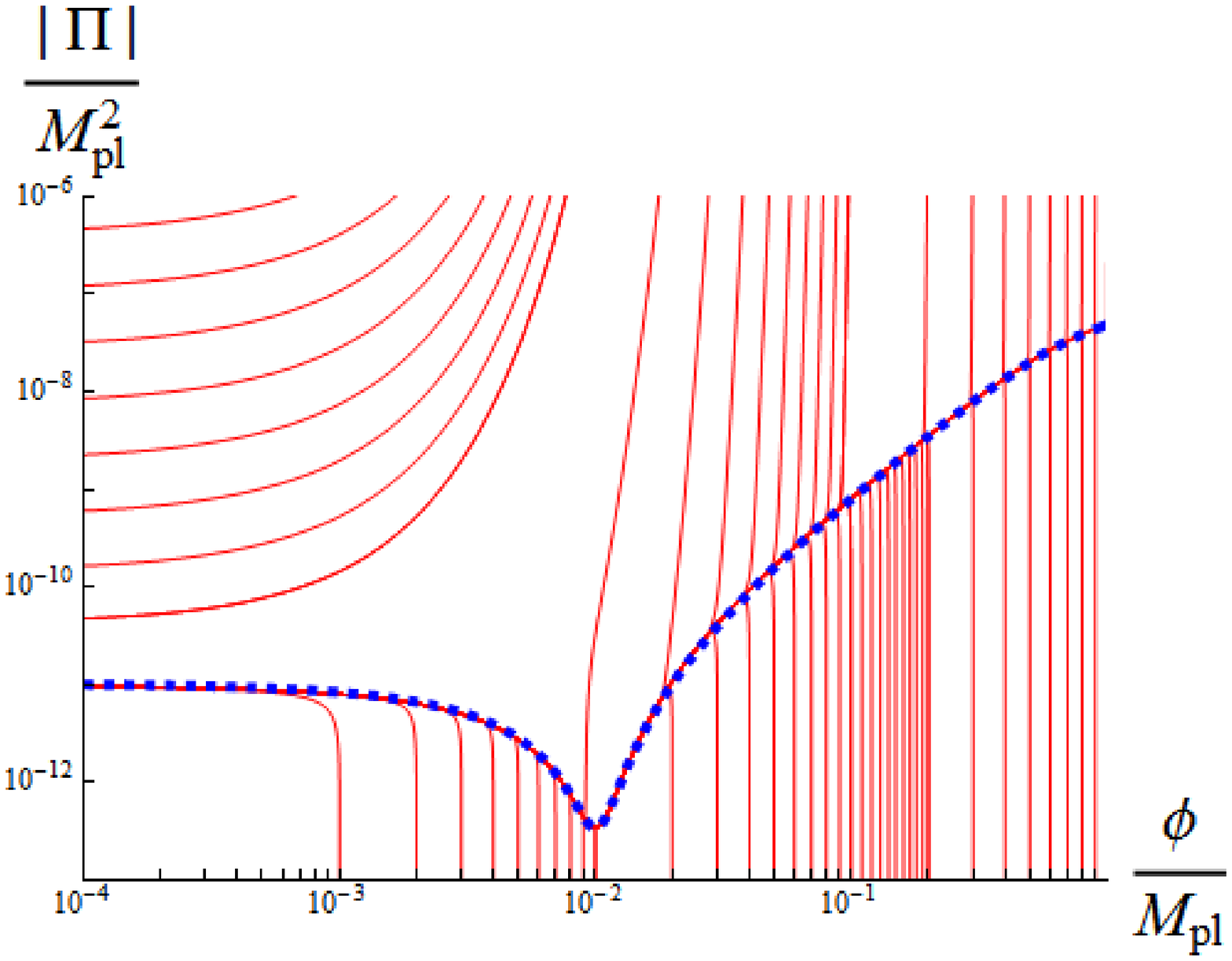} \includegraphics[scale=.48]{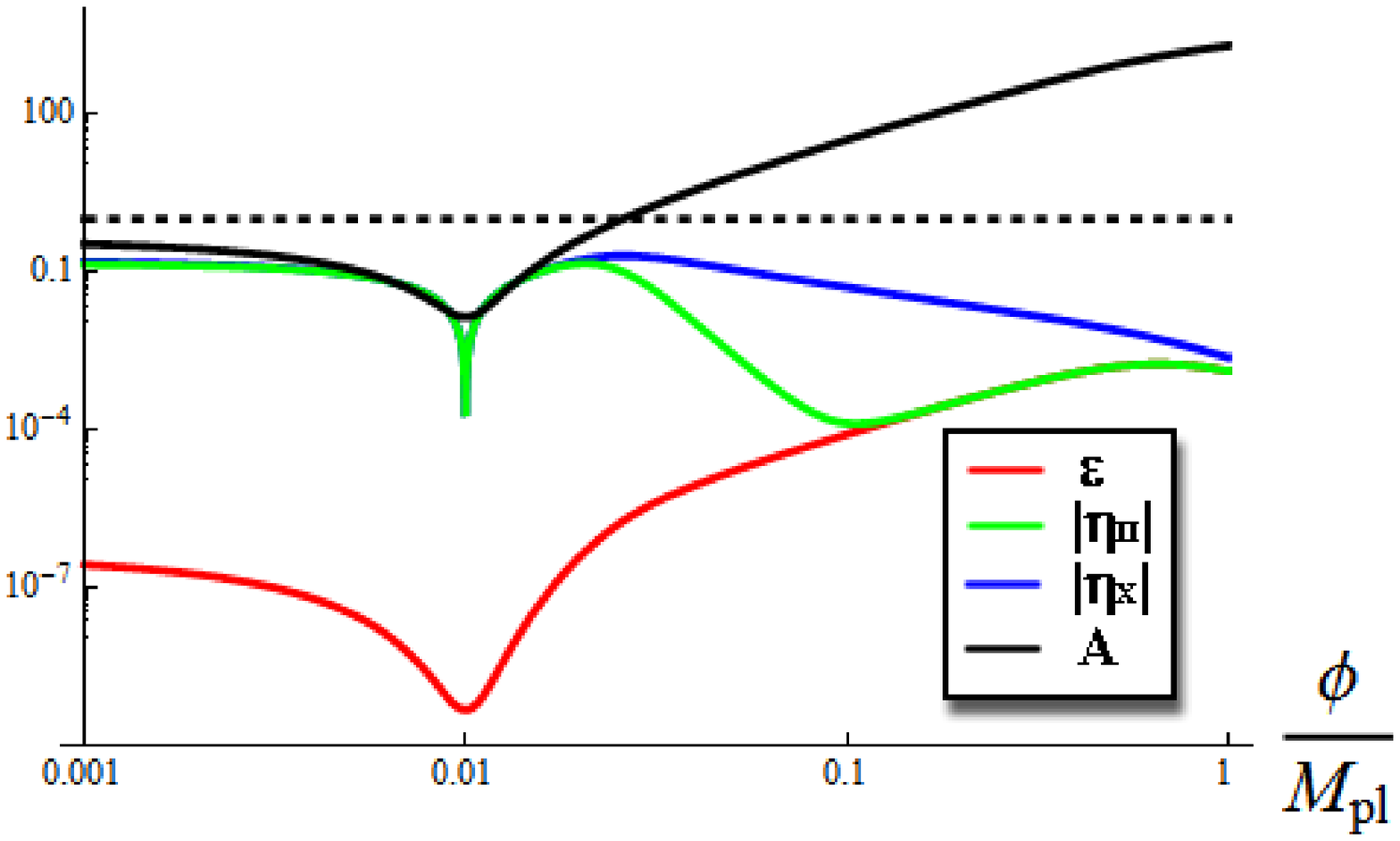}}
\caption{\small Left: The $(\phi,|\Pi|)$ phase-space diagram for a scalar field with a DBI kinetic term
and inflection-point potential (\ref{eq:inflectionpt}).  The thick dashed line is the inflationary attractor trajectory
and the thin solid lines are exact numerical solutions to the equations of motion.
Right: The inflationary parameters $\epsilon,|\eta_X|,|\eta_\Pi|$ and the
``noncanonical-ness" parameter $A(\phi)$ are shown as functions of $\phi$.  The dashed horizontal (black)
line denotes the value 1.}
\label{fig:DBIPhasePlot}
\end{figure}

\begin{figure}[htp]
\centerline{\includegraphics[scale=.38]{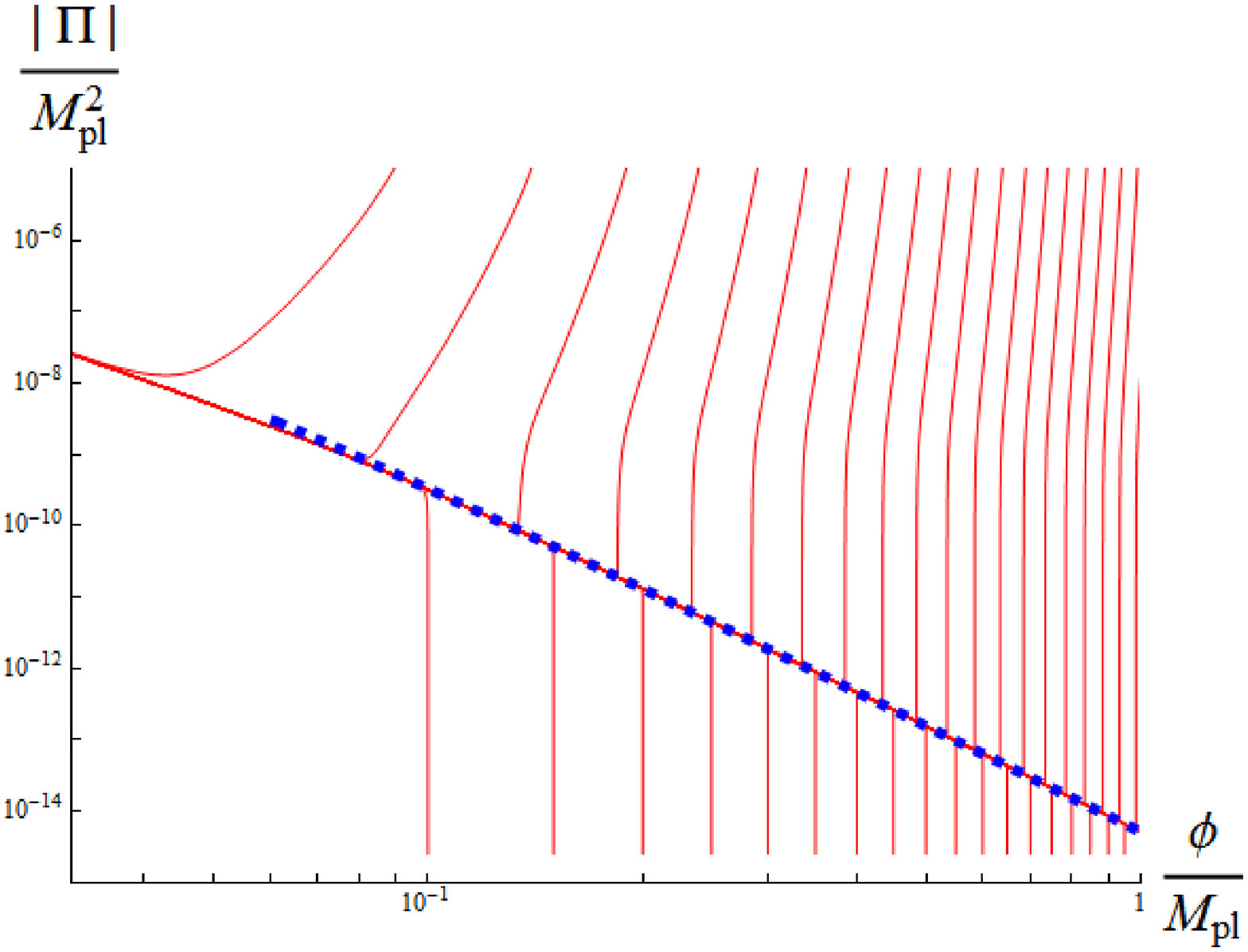} \includegraphics[scale=.50]{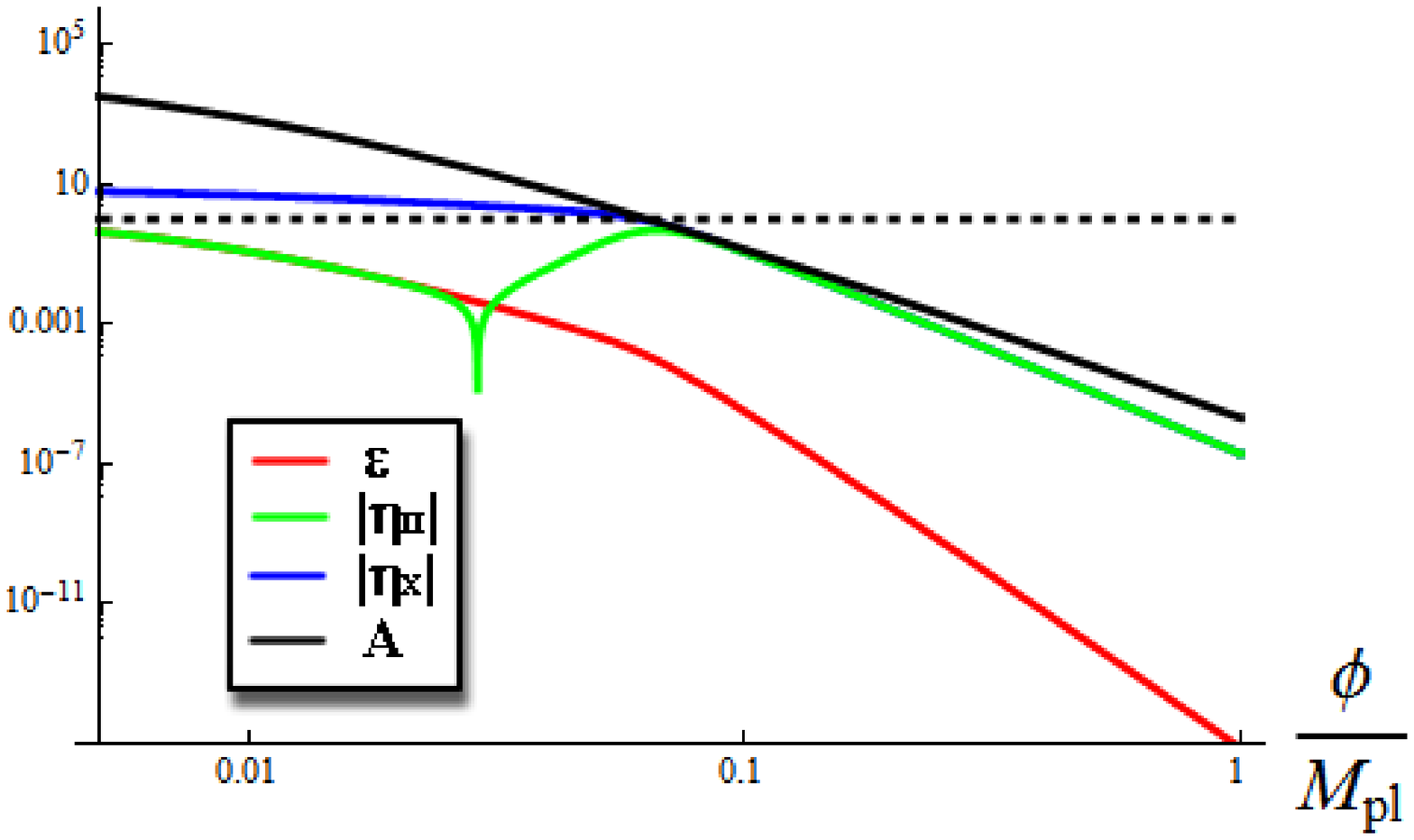}}
\caption{\small Left: The $(\phi,|\Pi|)$ phase-space diagram for a scalar field with a DBI kinetic term
and Coulomb potential (\ref{eq:coulomb}).  The thick dashed line is the inflationary attractor trajectory
and the thin solid lines are exact numerical solutions to the equations of motion.
Right: The inflationary parameters $\epsilon,|\eta_X|,|\eta_\Pi|$ and the
``noncanonical-ness" parameter $A(\phi)$ are shown as functions of $\phi$.  The dashed horizontal (black)
line denotes the value 1.}
\label{fig:DBI_coulomb_PhasePlot}
\end{figure}

As a concrete example, on the left-hand side of Figure \ref{fig:DBIPhasePlot} is the phase-space plot of the inflationary
solution, including the DBI regime, for a constant warp factor $f = (5\times 10^{-6} M_p)^{-4}$ and the same inflection-point
potential as above.  On the right-hand side of Figure \ref{fig:DBIPhasePlot}, the inflationary parameters and the parameter $A(\phi)$ are plotted as functions of $\phi$. There are several features that stand out
compared to the canonical
case in Figure \ref{fig:CanonPhasePlot}.  In particular, note that the inflationary parameters all stay below unity
for the entire range considered, courtesy of the DBI effect.  
Because of this, the inflationary attractor is valid
for a larger range of $\phi$ in the phase space.  Also, we see the $\eta_X \sim \epsilon$ behavior in
the large $A$ limit, as expected by our analysis (\ref{etaXdbi}), and that initial conditions
away from the inflationary trajectory
appear to be more strongly attracted to the attractor.  
Similar behavior is seen in Figure \ref{fig:DBI_coulomb_PhasePlot} for DBI inflation
on a Coulomb-like potential (\ref{eq:coulomb}) with warp factor $f=\left(5\times 10^{-5}M_p\right)^{-4}$
(the warp factor is different because the Coulomb potential has a different scale from that of the inflection-point potential, so the warp factor should change accordingly).  Note that usually DBI inflation
does not work on the Coulomb potentials that arise from brane inflation.  The reason we
have DBI inflation on a Coulomb potential here is that we have decoupled
the scales of $V(\phi)_{coulomb}$ and $f(\phi)$. In brane inflation models, these scales are set by the physics, and turn out to be such that DBI inflation on a Coulomb potential is impossible (see for example \cite{Bird:2009pq}).

\subsection{Tachyon Action}

Another popular Lagrangian that appears in string theory models of scalar fields has the form
\begin{equation}
p(X,\phi)_{Tach} = -V(\phi) \sqrt{1-2 \frac{X}{\Lambda^4}} \, .
\label{eq:Tachyon}
\end{equation}
Here $\Lambda$ is some constant scale (e.g.~the string scale), so that the
argument in the square root is dimensionless.
This type of Lagrangian most commonly appears as the effective Lagrangian for the open string tachyon
in bosonic string field theory describing the decay of an unstable D-brane
\cite{SenTachyon1,SenTachyon2,SenTachyon3,SenTachyon4}, but interestingly this Lagrangian also appears
(with a different function $V(\phi)$) as the effective action for a wrapped $D4$-brane in a
Nil-manifold as described in Eq(2.9) of \cite{BraneMonodromy}.
Inflation with the tachyon has been considered before in \cite{Cremades:2005ir,Raeymaekers:2004cu,LouisSarah};
we will only briefly consider this case, rederiving many of the results already known.
The conjugate momentum and energy density are
\begin{eqnarray}
\Pi_{Tach} &=& -\frac{V(\phi)}{\Lambda^4}\frac{\sqrt{2X}}{\sqrt{1-2X/\Lambda^4}} \\
\rho_{Tach} &=& \frac{V(\phi)}{\sqrt{1-2X/\Lambda^4}}\, .
\end{eqnarray}

In solving for the inflationary solutions, as with the DBI action above it is convenient to write quantities in terms
of the dimensionless parameter,
\begin{equation}
A_{Tach} \equiv \left(\frac{M_p^2}{3} \left(\frac{V'}{V}\right)^2 \frac{\Lambda^4}{V(\phi)}\right)^{1/2}
\end{equation}
which, as can be seen, is related to the usual canonical slow-roll parameter.
After some algebra, the solutions are found to be
\begin{eqnarray}
X_{inf}(\phi)   &=& \begin{cases}
    \frac{\Lambda^4 A_{Tach}^2}{2} & A_{Tach}\ll 1, \mbox{ ``slow roll" limit} \cr
    \frac{\Lambda^4}{2}\left(1-\frac{1}{A_{Tach}^{4/5}}\right) & A_{Tach} \gg 1 \mbox{ ``higher derivative" limit} \cr
    \end{cases}
\end{eqnarray}
The inflationary parameter becomes
\begin{equation}
\epsilon_{Tach} = \begin{cases}
  \frac{3}{2} A_{Tach}^{2} & A_{Tach} \ll 1 \cr
  \frac{3}{2} & A_{Tach} \gg 1\,.  \cr
  \end{cases}
\end{equation}
This form of Lagrangian does not support inflation in the limit in which the higher-dimensional operators
are important: Lagrangians of the type (\ref{eq:Tachyon}) only have the usual slow-roll-type solutions, irrespective
of the form of the ``potential" $V(\phi)$.  This result was in fact already noticed in
\cite{LouisSarah}, where it was found that the requirement $\rho + 3 p < 0$ for accelerated expansion to
occur requires $X/\Lambda^4 \leq 1/3$.  The Lagrangian satisfying this bound is approximately the canonical
one (\ref{eq:canon}), so inflationary solutions should only exist in the
canonical slow-roll limit, as we confirmed with our explicit inflationary analysis.

\subsection{Powerlike Lagrangian}

\begin{figure}[ht]
\centerline{\includegraphics[scale=.5]{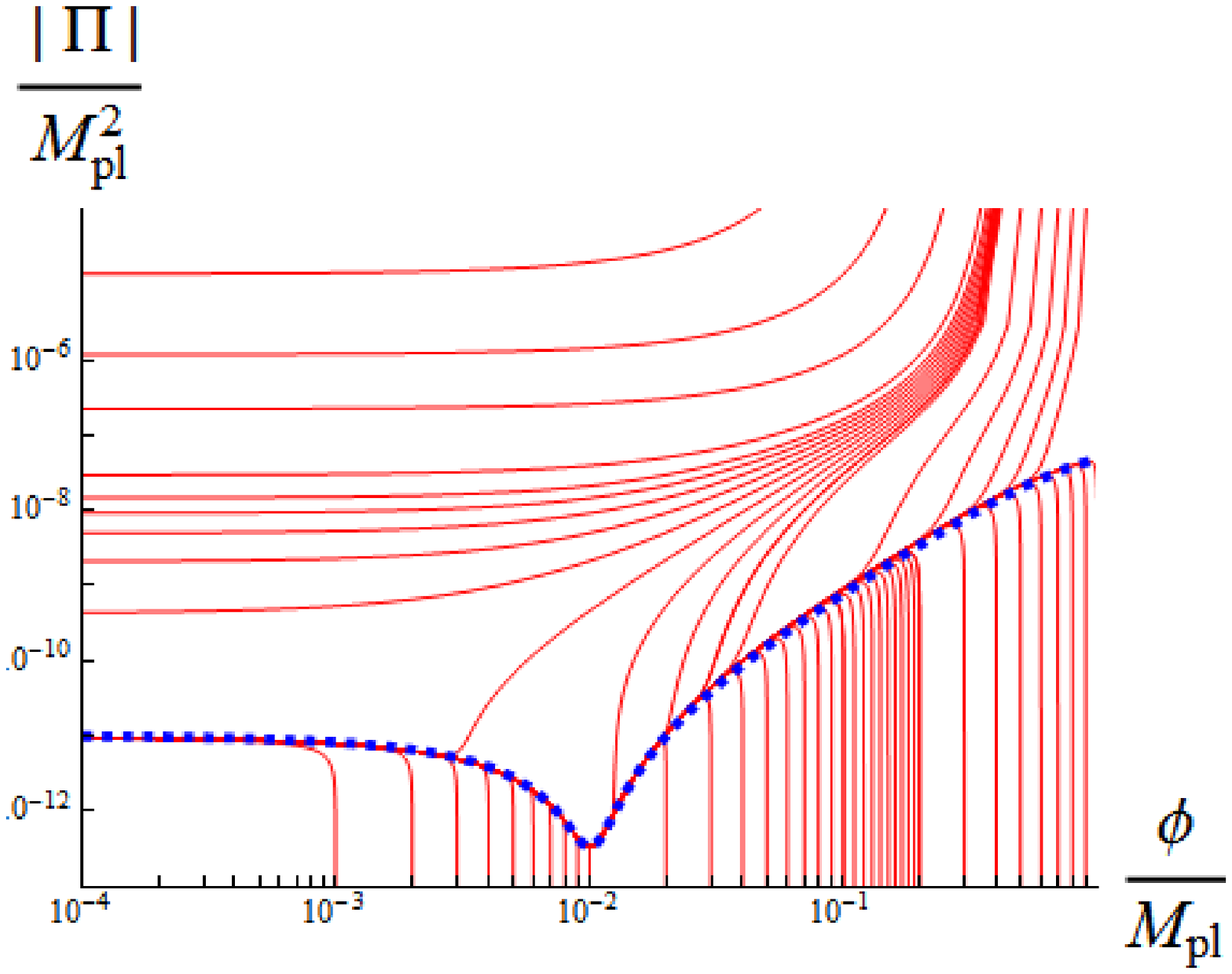} \includegraphics[scale=.48]{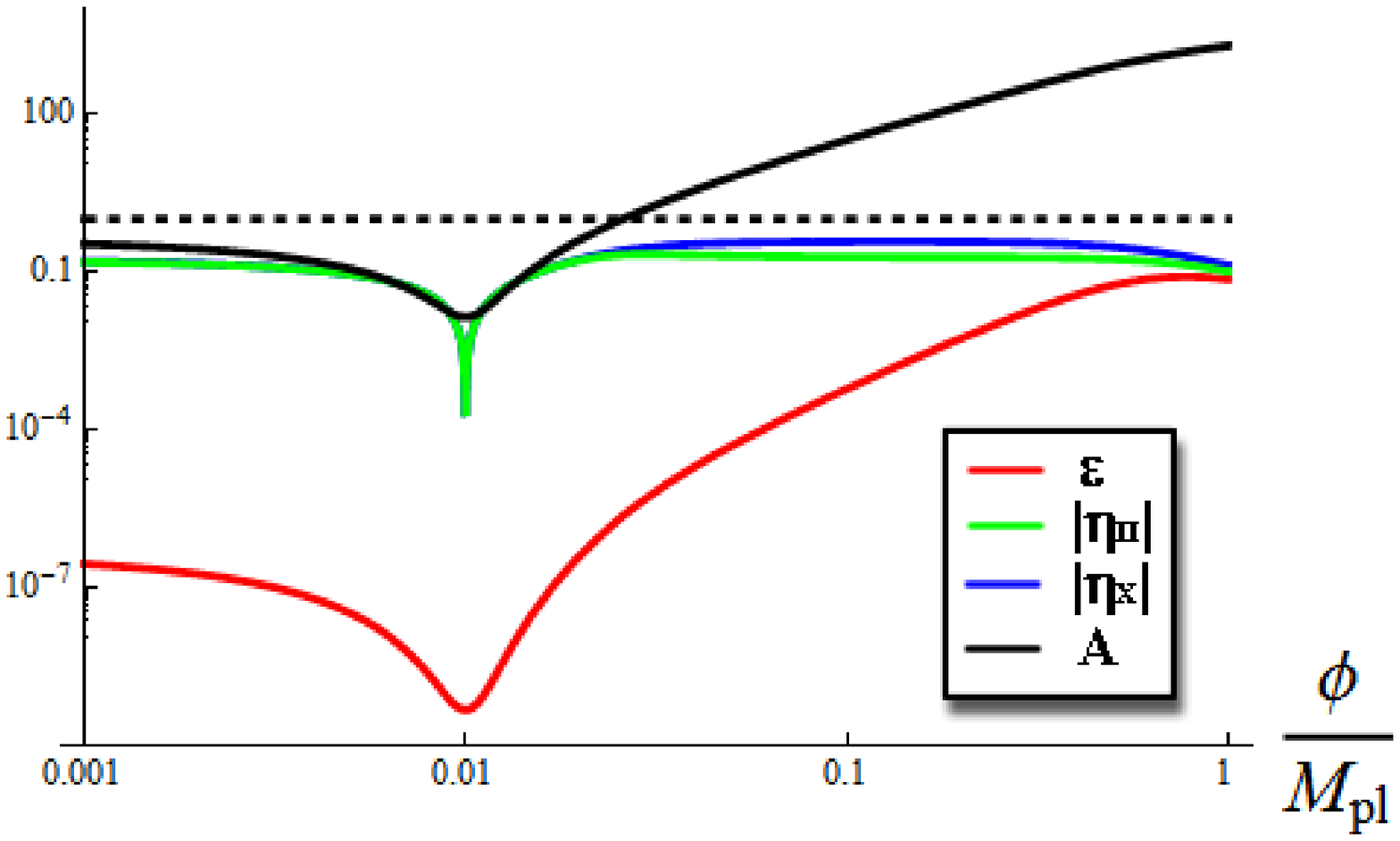}}
\caption{\small Left: The $(\phi,|\Pi|)$ phase-space diagram for a scalar field with a powerlike
kinetic term (\ref{eq:powerlikeEx})
and inflection-point potential (\ref{eq:inflectionpt}).  The thick dashed line is the inflationary attractor trajectory
and the thin solid lines are exact numerical solutions to the equations of motion.
Right: The inflationary parameters $\epsilon,|\eta_X|,|\eta_\Pi|$ and the
``noncanonical-ness" parameter $A(\phi)$ are shown as functions of $\phi$.  The dashed horizontal (black)
line denotes the value 1.}
\label{fig:PowerlikePhasePlot}
\end{figure}

\begin{figure}[ht]
\centerline{\includegraphics[scale=.45]{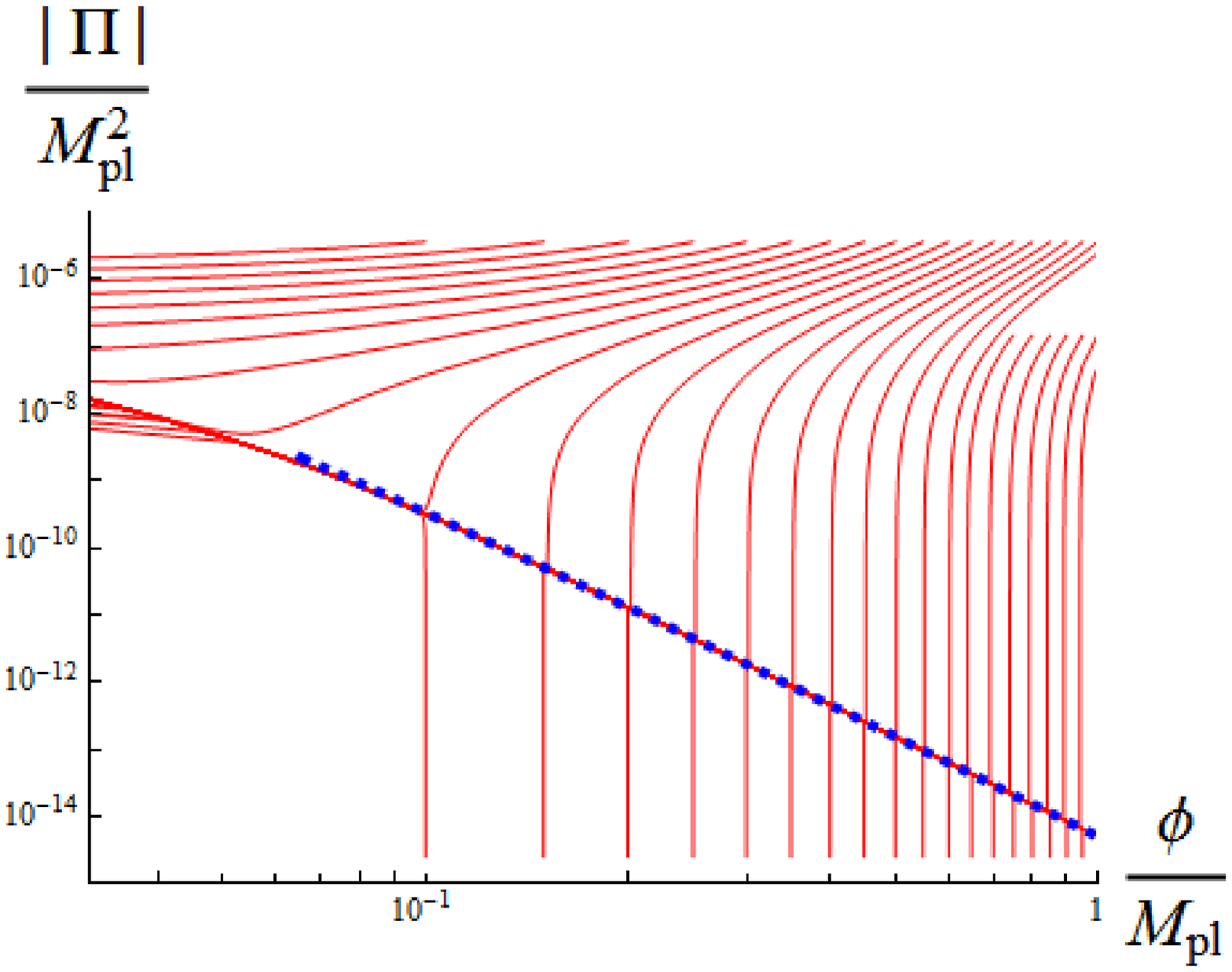} \includegraphics[scale=.48]{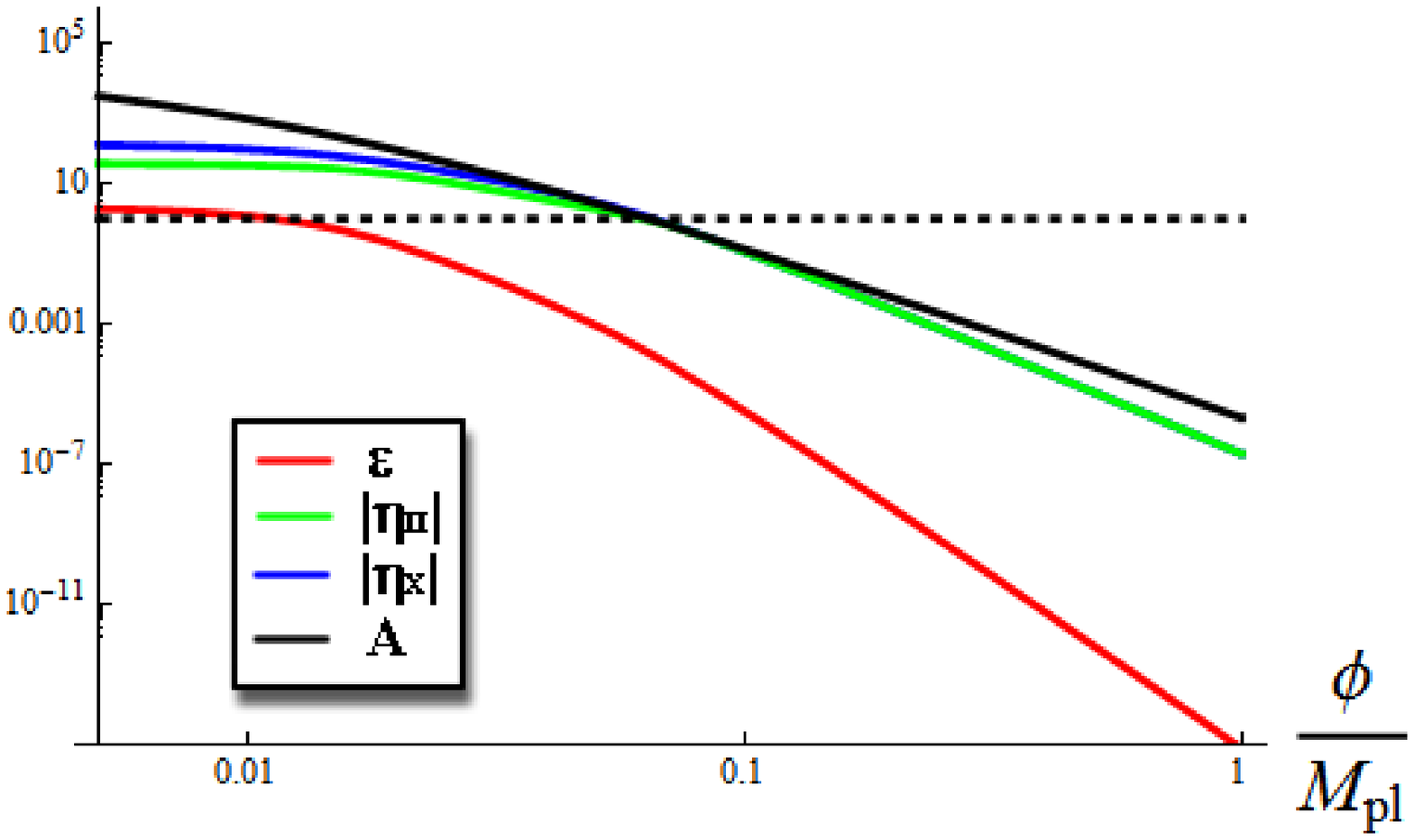}}
\caption{\small Left: The $(\phi,|\Pi|)$ phase-space diagram for a scalar field with a powerlike
kinetic term (\ref{eq:powerlikeEx})
and Coulomb potential (\ref{eq:coulomb}).  The thick dashed line is the inflationary attractor trajectory
and the thin solid lines are exact numerical solutions to the equations of motion.
Right: The inflationary parameters $\epsilon,|\eta_X|,|\eta_\Pi|$ and the
``noncanonical-ness" parameter $A(\phi)$ are shown as functions of $\phi$.  The dashed horizontal (black)
line denotes the value 1.}
\label{fig:Power_coulombPhasePlot}
\end{figure}

As an example of the general Lagrangians discussed in Section \ref{sec:GeneralAttractors},
let us take the kinetic term to be of the powerlike form:
\begin{equation}
p(X,\phi) = \Lambda^4 \left[\left(1+\frac{2}{3} \frac{X}{\Lambda^4}\right)^{3/2}-1\right] - V(\phi)\, .
\label{eq:powerlikeEx}
\end{equation}
For $X/\Lambda^4 \ll 1$, this Lagrangian has a canonical kinetic term plus corrections,
but it is very different at large $X$.  Note also that this Lagrangian does not
imply a speed limit; $X$ can take any value $X\in [0,\infty)$. Following the same analysis as in
the examples above, an inflationary solution can easily be found for all values of $A$:
\begin{equation}
X_{inf} = \frac{3}{4} \Lambda^4 \left[\sqrt{1+\frac{4}{3}A^2}-1\right]\, .
\end{equation}
The sound speed on the inflationary solution becomes
\begin{equation}
c_s^2 = \frac{1}{2} \left ( 1 + \frac{1}{\sqrt{1 + \frac{4}{3} A^2}} \right ) \, .
\end{equation}
Notice that $c_s^2 \geq 1/2$ even for $A\gg 1$, so there is a limit to how ``noncanonical"
this Lagrangian can become.
The inflationary parameters can be evaluated to be
\begin{eqnarray}
\epsilon &=&
  \frac{3}{2^{3/2}} \frac{\epsilon_{SR}}{A^2}\left(\sqrt{1+\frac{4}{3}A^2}-1\right)\left(1+\sqrt{1+\frac{4}{3}A^2}\right)^{1/2}
  \approx \begin{cases}
  \epsilon_{SR} & A \ll 1 \cr
  3^{1/4} \frac{\epsilon_{SR}}{A^{1/2}} & A \gg 1\,\cr
  \end{cases}; \\
\eta_X &\approx & \frac{3^{1/4}}{2}\frac{\epsilon_{SR}+\eta_{SR}}{A^{1/2}};\hspace{.6in} \eta_\Pi \approx  3^{1/4} \frac{\eta_{SR}}{A^{1/2}}\, ,
\end{eqnarray}
in the large A limit.
To demonstrate that this effect really is present, in Figures \ref{fig:PowerlikePhasePlot} and
\ref{fig:Power_coulombPhasePlot} we present
the phase-space plots of the inflationary solution and the inflationary parameters evaluated as a function
of $\phi$ for the same inflection-point potential (\ref{eq:inflectionpt}) and Coulomb potential (\ref{eq:coulomb})
as before, with $\Lambda = \left\{5\times 10^{-6}\, M_p,5\times 10^{-5}\right\}$ for the inflection-point and Coulomb potentials respectively,
i.e.~the same UV cutoff scales that were used in the DBI example above.  
By comparison with Figures \ref{fig:CanonPhasePlot}
and \ref{fig:CanonPhasePlot_Coulomb}
we see that the inflationary parameters are indeed suppressed relative to their usual canonical slow-roll values,
but not as much as for DBI inflation, as in Figures \ref{fig:DBIPhasePlot} and \ref{fig:DBI_coulomb_PhasePlot}.

\newpage

\section{Conclusion}
\label{sec:conclusion}

In this paper we have examined the effect of noncanonical kinetic terms on inflation in single-field models.
For effective Lagrangians of the form $p(X,\phi)$ (where $X = \frac{1}{2}\dot \phi^2$) we outlined how to
construct the general noncanonical inflationary solutions to the scalar field equations of motion.
The noncanonical inflationary solutions are
parameterized by the generalised inflationary parameters $\epsilon, \eta_X$ and $\eta_\Pi$, which smoothly
reduce to the usual slow-roll parameters in the canonical limit.
Furthermore, we have seen that inflationary trajectories are attractors, so that
deviations from an inflationary trajectory in phase space are driven to zero.
This analysis reduces to the known results for canonical slow-roll inflation, and can also be carried out
in the Hamilton-Jacobi formalism (see Appendix \ref{sec:HJ}), making the result quite general.

However, the fact that noncanonical inflation
is dynamically attractive obviously does not imply that all theories with noncanonical kinetic terms
are physically attractive.
In order for a model to give rise to noncanonical inflation at all
the Lagrangian must have the right functional form $p(X,\phi)$, and
the initial conditions should be such that inflation actually occurs.

In Section \ref{sec:GeneralAttractors} we described the
structure of Lagrangians that lead to noncanonical inflation.
We considered Lagrangians that are written as a power series expansion of the form
\begin{equation}
p(X,\phi) = \sum_{n\geq 0} c_n \frac{X^{n+1}}{\Lambda^{4n}} - V(\phi),
\label{series}
\end{equation}
with nonzero radius of convergence $R$, and as some known closed-form expression
\begin{equation}
p(X,\phi) = \Lambda^4 q\left(\frac{X}{\Lambda^4}\right) - V(\phi)\, ,
\label{closed}
\end{equation}
where in both cases $\Lambda$ is the UV cutoff of the effective theory.
For both (\ref{series}) and (\ref{closed}), noncanonical inflation is possible when both
$A\equiv \frac{V'}{3H}\frac{1}{\Lambda^2}$ and
$\frac{V}{\Lambda^4}$ are large.
For the series-expansion Lagrangian
(\ref{series}) to support noncanonical inflation, $\frac{\partial p}{\partial X}$ must further
diverge at $R$, while for the closed-form Lagrangian (\ref{closed}), the kinetic term must have a
positive second derivative with respect to X. This condition, $\frac{\partial^2 q}{\partial X^2} > 0$,
is just the condition that the speed of propagation of perturbations be real and subluminal, which
one should in any case demand of a physical theory.
In addition, we found that when a strong $\phi$ dependence in the kinetic
term is allowed, the inflationary parameter $\eta_X$ is generically large, and therefore that inflation
does not last for more than a few e-foldings.
(It should be noted that restricting to Lagrangians that have weak dependence on $\phi$ in
their kinetic terms may represent some fine-tuning from the effective-field-theory point
of view.)\footnote{We would like to thank L.~Leblond for discussions on this point.}
These results should be useful for model builders of inflation, regardless of the origin of the effective field theory
they are considering.

We examined these constraints for several examples, showing that - contrary to what one might have
guessed from known examples - the existence of a speed limit on $\phi$ is neither necessary
nor sufficient for the existence
of noncanonical inflationary solutions.
We explored the phase-space behaviour for inflationary attractors,
confirming that for noncanonical models the inflationary attractor is valid for a larger range of $\phi$ values
than in the canonical case.

It is also important to understand the set of
initial conditions that gives rise to (canonical or noncanonical) inflation. Trajectories with an
arbitrarily large initial momentum can miss the attractor trajectory altogether, because inflationary
attractors are typically only defined over some finite range of $\phi$ values. Theories with a severe overshoot
problem thus require very finely tuned initial conditions. Inflation in these models, albeit attractive, is not
especially natural. The presence of noncanonical terms can have an ameliorating effect on the severity of the
overshoot problem in a given theory, but their relevance in the allowed phase space is model specific,
as seen for DBI models in \cite{AttractiveBrane,Bird:2009pq}.
We are currently working on elucidating the role noncanonical kinetic terms play in the initial conditions
fine-tuning problem, and will return to
this subject in more detail in the near future \cite{Franche:2010yj}.

Although our effective -field-theoretic approach of studying the effect of higher-order kinetic terms in a
single scalar field theory can be self-consistent, it is also a great simplification. We have restricted ourselves
to the single-field case, whereas it is clear from the argument in the Introduction that the presence of other
fields, even those massive enough to be safely integrated out, can affect the inflationary
dynamics -- for instance via the noncanonical kinetic terms considered here. What is more, these fields are a
general feature of any attempt to embed an inflationary model in a UV-complete setting, such as string theory.
It would be interesting to understand better the effect of multiple fields in the context of our EFT approach.
We have also assumed
homogeneous initial conditions; however, since
inflation is invoked to explain observed large-scale homogeneity,
it would be helpful to understand how homogeneity can emerge dynamically from inhomogeneous initial conditions.
The presence of noncanonical kinetic terms can have an important and interesting dynamical
effect on this problem, and deserves further study.

\section*{Acknowledgments} We would like to thank Daniel Baumann,
Robert Brandenberger, Alejandra Castro, Xingang Chen, Jim Cline, Malcolm Fairbairn, Louis Leblond, Alexander Maloney, Nikolaos Mavromatos, Enrico Pajer, Hiranya Peiris, Mairi Sakellariadou, Sarah Shandera, Navin Sivanandam, Marcus Tassler, Henry Tye, and Wang Yi
for helpful discussions.
B.U.\ is supported in part through an IPP
(Institute of Particle Physics, Canada) Postdoctoral Fellowship, and
by a Lorne Trottier Fellowship at McGill University and would like to thank the
Aspen Center for Physics for hospitality while part of this work was completed.
The work of P.F. is supported by the Natural Sciences and Engineering Research Council (NSERC) of Canada. The work of A.W. is supported by the Fonds Qu\'eb\'ecois de la Recherche sur la Nature et les Technologies (FQRNT).R.G.~is supported by an NSERC Postdoctoral Fellowship, the STFC grant ST/G000476/1, and a Canada-UK Millennium Research Award.

\appendix



\section{Attractors in the Hamilton-Jacobi Formalism}
\label{sec:HJ}

We will now show that noncanonical inflation is an attractor using the Hamilton-Jacobi formalism.
The Hamilton-Jacobi equation for the general action $p(X,\phi)$ is (see Eq.(38) of \cite{Bean})
\begin{equation}
3 M_p^2 H^2(\phi) = \frac{4 M_p^4 H'^2}{p_X(X(H'),\phi)}-p(X(H'),\phi),
\label{eq:HJ}
\end{equation}
where a prime $'$ denotes a total derivative with respect to $\phi$ and
$X=X(H')$ is defined by the relation (see Eq.(35) of \cite{Bean})
\begin{equation}
H'(\phi) = \frac{\sqrt{2X}p_X}{2 M_p^2}\, .
\label{eq:Hprime}
\end{equation}
Note that
\begin{equation}
c_s^2 = \left(1+2 X\frac{p_{XX}}{p_X}\right)^{-1}\, .
\end{equation}

We want to consider perturbations
$H(\phi) = H_0(\phi)(1 + \delta H(\phi))$ about
a background solution $H_0(\phi)$, and are interested in scenarios when this
perturbation decays rapidly.\footnote{Notice that this ansatz for the perturbation
is different from the one that is usually taken, $H = H_0 + \delta \tilde H$.  
The difference between the two ans\"atze is proportional to the time variation
of the background Hubble parameter $H_0$, and indeed we will see below that our general
result only differs from the standard result up to terms of order the inflationary parameter $\epsilon$.
However, our improved perturbation ansatz removes the possibility of an
incorrect identification of an attractor,
such as when $\delta \tilde H$ decreases at the same rate as $H_0$. In this case our definition explicitly reveals the true perturbation, $\delta H(\phi)$, to be a constant, i.e. we do not have an attractor.}  
We will see that perturbations decay when $H_0(\phi)$ is
inflationary, but for now we take $H_0(\phi)$ to be general.
At the linearized level, we have
\begin{eqnarray}
H^2(\phi)&\cong& H_0^2 + 2H_0^2\delta H;\\
H'(\phi)&\equiv&H'_0+(H'_0\delta H+H_0\delta H');\\
H'^2(\phi)&\cong&(H'_0)^2+2H'_0(H'_0\delta H+H_0\delta H')+O(\delta H)^2.
\end{eqnarray}
It is useful to write the perturbed version of (\ref{eq:Hprime}) (at fixed $\phi$) as
\begin{eqnarray}
&& (H'_0\delta H+H_0\delta H')=-\delta X \frac{p_X}{2M_p^2\sqrt{2X}}c_s^{-2} \\
&& \Rightarrow \frac{\delta X}{(H'_0\delta H+H_0\delta H')}=\frac{4M_p^4c_s^2H'(\phi)}{(p_X)^2},
\label{eq:chainrule}
\end{eqnarray}
using the unperturbed version of (\ref{eq:Hprime}) to rewrite the right-hand side.

The linearized form of (\ref{eq:HJ}) under perturbations takes the following form (using the chain rule):
\begin{eqnarray}
6M_p^2H_0^2\delta H(\phi)&=&\frac{8M_p^4H'_0(H'_0\delta H+H_0\delta H')}{p_X} \nonumber \\
&-&\frac{4M_p^4(H'_0)^2p_{XX}}{(p_x)^2}\left(\frac{\delta X}{(H'_0\delta H+H_0\delta H')}\right)(H'_0\delta H+H_0\delta H')\ \ \ \ \ \ \ \nonumber \\
&-&p_{X}\left(\frac{\delta X}{(H'_0\delta H+H_0\delta H')}\right)(H'_0\delta H+H_0\delta H')
\label{eq:inter1}
\end{eqnarray}
Using (\ref{eq:chainrule}), (\ref{eq:inter1}) can be rewritten as
\begin{equation}
\frac{\delta H'}{\delta H}=-\frac{3}{2}\frac{H_0}{H'_0M_p^2}\left[\frac{c_s^2-2}{p_X}+4M_p^4c_s^2(H'_0)^2\frac{p_{XX}}{p_X^4}\right]^{-1}
  -\left(\frac{H'_0}{H_0}\right).
\label{eq:inter2}
\end{equation}
Making use of (\ref{eq:Hprime}) for $H' = H_0'$, we can rewrite the last term in the parentheses of (\ref{eq:inter2})
to be $c_s^2\ 2X p_{XX}/(p_X)^2$, so that (\ref{eq:inter2}) becomes the simple expression (after cancellation of terms involving $c_s$):
\begin{equation}
\frac{\delta H'}{\delta H} = \frac{3}{2\ M_p^2} \left(\frac{H_0}{H_0'}\right) p_X -\left(\frac{H'_0}{H_0}\right)\, .
\label{eq:HJpert}
\end{equation}

Now by the definition of the number of e-folds $N_e = \int H_0 dt$, we have
\begin{equation}
dN_e = H_0 dt = H_0 \frac{dt}{d\phi} d\phi = -\frac{H_0}{\sqrt{2X}} d\phi = - \frac{1}{2 M_p^2}\left(\frac{H_0}{H'_0}\right) p_X d\phi\, ,
\end{equation}
where we used (\ref{eq:Hprime}) to rewrite $\sqrt{2X}$.
This can be used to rewrite (\ref{eq:HJpert}) as
\begin{equation}\label{eqndiff}
\frac{d \ln \delta H}{dN_e} = -3 + \epsilon,
\end{equation}
where the Hamilton-Jacobi form for the inflationary parameter is \cite{Bean}
\begin{equation}
\epsilon = \frac{2M_p^2}{p_X} \left(\frac{H'}{H}\right)^2\, .
\end{equation}
The solution to (\ref{eqndiff}) is
\begin{equation}
\delta H\sim e^{(-3+\epsilon)N_e}.
\end{equation}
When the inflationary parameter is small, i.e. $\epsilon \ll 1$, we find as in \cite{HJ}
that perturbations $\delta H$ decay as
\begin{equation}
\delta H \sim e^{-3 N_e}\, .
\end{equation}
Since the number of e-folds grows rapidly during inflation, perturbations from noncanonical
inflationary solutions are
seen to be attractors in the Hamilton-Jacobi formalism.

\section{K-inflation Lagrangians}
\label{app:kinflation}
Our analysis in Section \ref{sec:GeneralAttractors} was not entirely general - the solutions we constructed assumed that the coefficients
of the higher-dimensional operators were positive.  When the coefficients are negative, one can obtain new
types of solutions with different behavior. These are the {\it k-inflationary}-type solutions studied
in \cite{kinflation}.  However, as discussed earlier, these types of Lagrangians typically violate, somewhere
in phase space, either
the null-energy condition or the conditions that the ``sound speed", $c_s$, of perturbations is less
than that of light, and real. Nevertheless, it is interesting to consider these types of Lagrangians
for completeness.

Let us consider a Lagrangian as in \cite{kinflation}, given by
\begin{equation}
p(X,\phi)_K = K(\phi) X + \frac{1}{\Lambda^4} X^2,
\end{equation}
where $K(\phi) < 0$ for some region of $\phi$.
In this case we find that the conjugate momentum is
\begin{equation}
\Pi_K = -\sqrt{2 X} \left[K(\phi) + \frac{2}{\Lambda^4} X\right]\, .
\end{equation}
Solving (\ref{eq:PiAtrSoln}) for $X$ with the above Lagrangian and canonical
momentum gives the solution
\begin{equation}
\label{eq:Xatrk}
X_{inf}^K = \frac{\Lambda^4}{2} (-K(\phi)) \left(1-\frac{1}{\sqrt{3}} \frac{M_p K'}{(-K)^{3/2}}\right)
+ {\mathcal O}\left(\frac{M_p K'}{(-K)^{3/2}}\right)\, .
\end{equation}
Note that this matches the solution found in \cite{kinflation}(Eq.(4.6)) to leading order; here
we can compute the subleading corrections as well.

Using (\ref{eq:Xatrk}) we can compute the inflationary parameters:
\begin{eqnarray}
\epsilon &=& 2\sqrt{3} M_p \frac{K'}{(-K)^{3/2}}; \\
\eta_\Pi &=&  2\sqrt{3} \frac{M_p}{(-K)^{1/2}} \frac{K''}{K'}; \\
\eta_X &=& \epsilon+\sqrt{3} M_p \frac{K'}{(-K)^{3/2}} = \frac{3}{2} \epsilon\, .
\end{eqnarray}
We see that $\epsilon_K$ matches the small parameter $\delta X/X_{sr}$ found in \cite{kinflation}.
In addition, we find another parameter which must be small, proportional to
$\frac{M_p}{(-K)^{1/2}} \frac{K''}{K'}$, which would not have been seen with the analysis
of \cite{kinflation}.  

\section{DBI Inflation ${\mathcal O}(\Delta)$ Terms}
\label{sec:DBIDelta}

In this section we revisit the assumption made for the DBI system that DBI inflationary solutions only
exist for $|\Delta| \ll 1$.
The general solution to (\ref{eq:chifixedpt}) is not immediately apparent. However, the solution for the case $\Delta = 0$ is easily found, and perturbing this solution gives a series expansion in $\Delta$. As argued below, we have several reasons to expect $\Delta$ to be small, making this expansion valid. Here we assume that the energy density is dominated by the potential so that  $H(\phi)  = V^{1/2}(\phi)/{\sqrt{3}M_p}$ is a function of $\phi$; deviations to this will be proportional to the slow-roll parameters. We are then able to treat $A$ as a constant in $X$ when solving (\ref{eq:chifixedpt}): $A = \left(\frac{2}{3} \epsilon_{SR} V f\right)^{1/2}$. This assumption is motivated by the fact that it is required for inflation to take place; its consistency is also checked below. The solution to first order in $\Delta$ is
\begin{eqnarray}
\label{eq:XDBIApp}
X_{inf}^{DBI}(\phi) & = &\frac{ A^2Q^2}{2f} - \frac{\Delta A Q^3}{f}(1 - Q - \frac{1}{2}A^2 Q^2);\\
\Pi_{inf}^{DBI}(\phi) &=&  -\frac{M_p V'}{\sqrt{3 V}} + \frac{\Delta}{\sqrt{f} Q}(1 - Q - \frac{1}{2}A^2 Q^2);\\
\rho_{DBI} &=& \frac{1}{fQ} - \frac{1}{f} +  V  - \frac{\Delta A}{f} (1 - Q - \frac{1}{2}A^2 Q^2)\, ,
\label{eq:DBIrho}
\end{eqnarray}
where we defined $Q = \frac{1}{\sqrt{1+A^2}}$.

With this solution in hand, we can now check our assumption that the energy
density is dominated by the potential energy.  It is clear that for $A \ll 1 \Rightarrow Q \approx 1$, the energy density (\ref{eq:DBIrho}) is dominated by the potential energy $V$.
In the limit where $A \gg 1$, the scalar field approaches the speed limit $X_{max} = \frac{1}{2f}$, as can be seen from
(\ref{eq:XDBIApp}).  This might be a concern, since the energy density,
\begin{equation}
\rho = \frac{1}{f(\phi)} \left[ \frac{1}{\sqrt{1-X/X_{max}}}-1\right] + V(\phi),
\end{equation}
appears to be dominated by the kinetic term in this limit.  However, we see that the energy density
evaluated on the DBI inflationary attractor solution given above,
\begin{equation}
\rho_{DBI} \approx V \left[ 1+ \frac{A}{V f}\right] = V \left[1+\sqrt{\frac{2}{3}} \sqrt{\frac{\epsilon_{SR}}{Vf}}\right]\, ,
\end{equation}
is still dominated by the potential as long as $V f \gg A$.
This conclusion does not change when the ${\cal{O}}(\Delta)$ terms are included: $\rho \approx V \left (1 + \frac{(1-\Delta)A}{Vf} \right )$ is still dominated by $V$ for $Vf \gg A$. Thus when $\epsilon_{SR} \sim {\mathcal O}(1)$ (i.e.~the potential is not flat), our assumption about the energy density
- and of the existence of the inflationary solution - is only valid when $V \gg 1/f$. For example, the potential must be large in units of the ``warped" energy scale
$\Lambda_{warped} = 1/f(\phi)^{1/4}$.

The ${\cal{O}}(\Delta_{DBI})$ corrections to the DBI inflationary parameters can be calculated and are given (in terms of Q) by
\begin{eqnarray}
\epsilon_{DBI} & = & \frac{3A^2Q^2}{2(1 + Q(fV-1))} + \frac{3AQ^5 \Delta ((2+A^2)(fV-1)Q +2)(2-(2+A^2)Q)}{4(1+(fV-1)Q)^2};\\
(\eta_X)_{DBI} &=& \epsilon_{DBI} + \frac{1}{(A^2 - 2\Delta AQG)^{\frac{1}{2}}}\times \left   [\left ( \eta_{SR} - \epsilon_{SR}\right ) Q^3 A \right.
 \\ \nonumber && - \Delta \left ( \eta_{SR} - \epsilon_{SR} \right) Q^4 \left(G(1-2A^2) + A^2Q(1-Q) \right)   + \frac{\Delta A^2 Q}{2} \left ( \frac{GQ}{f M_p^2 H^2} + 3(Q^2 - 1) \right ) \\ \nonumber && \left .+ \frac{3 \Delta^2 A Q^2}{2} \left ( 5G - Q^2(G(1-2A^2) + A^2Q(1-Q)) \right )  - \Delta^{(2)}AGQ^2 \right ];\\
(\eta_\Pi)_{DBI} &=& - \frac{\sqrt{1-M}}{2M ( - \Delta (1- \sqrt{M})^2 + 2A \sqrt{M})}\times \\\nonumber && \left [-4 \eta M^{3/2} + 2 M \Delta^{(2)} (1-\sqrt{M})^2 + 3 \Delta^2 (1-6M - 3M^2 + 8M^{3/2}) \right],
\end{eqnarray}
where for compactness we have defined $G = 1 - Q - \frac{1}{2} A^2 Q^2$, $M = 1-A^2Q^2 + 2 \Delta A Q^3 G$ and
the parameter $\Delta^{(2)} \equiv M_p^2 f''/(V f^2)$. At this point we can check our earlier assumptions
that $\Delta_{DBI}$ is small, and expanding in terms of it is valid. To simplify matters we work in the DBI
limit where $A$ is large; in this limit the inflationary parameters, including ${\mathcal O}(\Delta)$
corrections, are given by
\begin{eqnarray}
\left . \epsilon_{DBI} \right|_{A \gg 1} & = & \frac{3 A}{2 f V} \left(1-\frac{1}{2}\frac{\Delta}{A}\right);\\
\label{etalargeA} \left .(\eta_X)_{DBI} \right |_{A \gg 1} &=& \epsilon_{DBI} + \frac{(\eta_{SR} - \epsilon_{SR})}{A^3}\left(1+\Delta\right) - \frac{3}{2} \Delta
- \frac{\Delta^{(2)}}{2A};\\
\label{etapilargeA} \left. (\eta_\Pi)_{DBI} \right |_{A \gg 1} &\approx& \frac{\eta_{SR}}{A} \left(1+\frac{1}{2}\Delta^{1/2}\right)- \frac{\Delta^{(2)}}{2 \Delta^{1/2}} - \frac{\Delta^{(2)}}{4} - \frac{3}{4} \Delta^{1/2}A^2.
\end{eqnarray}
Clearly, since $\eta_X \sim \Delta$, we will require $\Delta \ll 1$ in order to have inflationary solutions.
However, since $\eta_\Pi \sim \Delta^{1/2} A^2$, the smallness of $\eta_\Pi$ requires the even
stronger condition that $\Delta \ll 1/A^4$.  We also see that the parameter $\Delta^{(2)}$ is constrained
to be small; the strongest constraint again comes from $\eta_\Pi$, which requires $\Delta^{(2)} \ll \Delta^{1/2} \ll 1/A^2$.
Thus, our earlier assumption that $\Delta$ is small is consistent.
Similarly, the requirement that the sound speed $c_s^2$ (\ref{eq:SoundSpeddDef}) needs to be
approximately constant during inflation is imposed through the smallness of the additional inflationary
parameter $\kappa$:
\begin{equation}
\kappa \equiv \frac{\dot{c_s}}{Hc_s} = -\frac{1}{2} \frac{1}{1+\frac{p_X}{2Xp_{XX}}}
  \left(\frac{\dot X}{HX} + \frac{\dot p_{XX}}{H p_{XX}} - \frac{\dot p_{X}}{H p_{X}}\right).
\end{equation}
For the DBI Lagrangian on the inflationary solution this gives
\begin{eqnarray}
\kappa_{DBI} =&& \frac{1}{2} \frac{2 f X_{inf}}{1-2 f X_{inf}} \left[3\Delta \sqrt{2 f X_{inf}} - 2 (\epsilon-\eta_X)\right] \nonumber\\
  && \stackrel{A\gg 1}{\rightarrow} \left(\frac{\eta_{SR}-\epsilon_{SR}}{2A}\right)(1+\Delta) + 3 A^2 \Delta + A \Delta^{(2)}\,.
\end{eqnarray}
The above requirements of having small $\Delta$ and $\Delta^{(2)}$ are sufficient to keep $|\kappa_{DBI}| \ll 1$, as long
as we are in an inflationary regime where $\epsilon_{SR}/A$ and $\eta_{SR}/A \ll 1$.

As an example where $\Delta$ is small, in the original string theory inspired models of DBI inflation \cite{DBI} the potential and warp
factors are $V(\phi) = m^2 \phi^2/2$, and $f(\phi) = \lambda/\phi^4$.
We obtain from (\ref{eq:epsilonattr}) $\epsilon_{DBI} \approx \sqrt{6/\lambda} M_p/m$, which agrees with the leading order behavior given in (3.8)
of \cite{ShanderaTye}.  For this warp factor, we have $\Delta_{DBI} \sim 1/(m \lambda^{1/2}) \sim \epsilon_{DBI}$, so the smallness
of $\Delta_{DBI}$ arises from the smallness of $\epsilon$, even though $f(\phi)$ is not actually constant in $\phi$.

\bibliography{kAttractors}

\end{document}